\documentclass[aps,prx,preprint,superscriptaddress,floatfix]{revtex4-2}

\usepackage{amsmath,amssymb,graphicx,booktabs,dcolumn,siunitx}
\usepackage{float}
\usepackage{color}
\usepackage{multirow}
\usepackage{booktabs}
\usepackage{graphicx}

\begin{document}

\title{Hierarchical Response Theory of Electronegativity from Atomic Mean Inner Potential}

\author{Jin-Cheng Zheng}
\affiliation{Department of Physics and China-Malaysia International Institute For Advanced Studies, Xiamen University, Xiamen 361005, China }
\affiliation{Department of Physics, Xiamen University Malaysia, Sepang 43900, Malaysia}
\email{jczheng@xmu.edu.my}

\date{\today}

\begin{abstract}
Electronegativity, introduced by Pauling nearly a century ago, remains
chemistry's most widely used yet conceptually ambiguous quantity: is it
a fixed atomic constant, or does it depend on chemical environment? We
resolve this debate by recasting electronegativity as a continuous
response function rooted in the Atomic Mean Inner Potential (AMIP), a
quantum-mechanical observable accessible through first-principles
calculation and, ultimately, electron holography. Building on the
established static AMIP scale \(\chi_{\mathrm{AMIP},1/2}\), we promote
it to a function \(\chi(\delta)\) of fractional positive charge
\(\delta\) and expand it in a Taylor series. The zeroth-order term
recovers the static scale; the first-order term defines the
Electrostatic Rigidity \(\mathcal{R}\), whose sign indicates whether an
atom hardens or softens upon oxidation; and the second-order term
defines the Electrostatic Anharmonic Modulus \(\mathcal{A}\). A
logarithmic-derivative decomposition separates \(\mathcal{R}\) into an
intrinsic, scale-independent response asymmetry
\(\mathcal{D}\equiv \mathcal{R}_\Omega/\Omega_0-\mathcal{R}_f/f_0\)
and the static scale \(\chi_0\), with
\(\mathcal{R}^{(p)}=c_p\,\chi_0^{(p)}\mathcal{D}\) and \(c_{1/2}=1/2\).
The dimensionless quantity \(\mathcal{D}\), which is independent of the
scale exponent \(p\), governs the competition between electronic
screening loss (softening) and geometric volume collapse (hardening)
for all 102 elements. This competition produces a systematic dichotomy
between softening main-group elements and hardening d-block elements,
two scaling laws within the transition series (closed-shell stiffening
and principal quantum softening), and two anomalies: a pronounced
exchange-driven effect at half-filled d\(^{5}\) configurations and a
near-perfect cancellation in Ce. It also resolves the long-standing
Cs--Fr controversy, identifying Fr as a critical element whose
near-boundary response is amplified fourfold by relativistic
contraction of the 7s orbital. The framework reproduces the classical
Allen, Pauling, and Mulliken scales as well as Lewis acidity trends,
and makes three falsifiable experimental predictions. It transforms
electronegativity from a lookup-table number into a predictive,
hierarchically structured quantum descriptor.
\end{abstract}

\keywords{Electronegativity, Atomic Mean Inner Potential, Fractional
Charge, Electrostatic Rigidity, Intrinsic Response Asymmetry, Chemical
Hardness, Density Functional Theory, Response Theory}

\maketitle

\section{Introduction}
\label{sec:intro}

Electronegativity stands among the most successful yet conceptually
unsettled concepts in chemistry. Since Pauling's original definition---
``the power of an atom in a molecule to attract electrons to
itself''~\cite{Pauling1932}---it has become the organizing principle
behind bond polarity, reaction trends, and the periodic table itself.
Yet the very success of the concept has obscured a fundamental
ambiguity: is electronegativity a fixed atomic constant, or does it
depend on the chemical environment in which the atom finds itself?
Most widely used scales assign a single number to each element, even
though it has long been recognized that electronegativity increases
with oxidation state, decreases with coordination number, and changes
with hybridization~\cite{Mulliken1934,walsh1948structures,LiXue2006}.
This tension has motivated a proliferation of context-specific
scales~\cite{DongOganov2022}, each capturing a particular facet of the
problem but none providing a unified physical foundation.

The difficulty is not merely empirical. Conceptual density functional
theory (CDFT) identifies electronegativity with the negative of the
electronic chemical potential, $\mu = \partial E/\partial N$, and
chemical hardness with its second derivative,
$\eta = \frac{1}{2}\partial^2 E/\partial N^2$~\cite{ParrDonnellyLevyPalke1978,ParrPearson1983}.
This framework correctly recognizes electronegativity as a response
function rather than a static property, but defines the response
through derivatives of the \textit{total energy}---quantities that are
inherently indirect and difficult to isolate experimentally. What is
missing is a formulation in which the response of the electron cloud
to charge perturbation is expressed in terms of a quantity that is
both physically transparent and directly measurable.

The atomic mean inner potential (AMIP)~\cite{Zheng2026AMIP} provides such a quantity.
Defined as the volume-averaged electrostatic potential experienced by
an electron inside an atom, the AMIP is accessible both through
first-principles calculation and through quantitative electron
holography~\cite{Gajdardziska1993,Mkhoyan2008}, and forms the basis of
our recently established static electronegativity
scale~\cite{Zheng2026AMIP}. That scale, expressed in its
square-root form \(\chi_{\mathrm{AMIP},1/2}\), reproduces the Allen
spectroscopic scale~\cite{Allen1989} with $R^2 = 0.97$, the Pauling thermochemical
scale~\cite{Pauling1932} with $R^2 = 0.88$ (rising to $0.94$ for main-group
elements), and the Mulliken ionization-energy-based scale~\cite{Mulliken1934} with
$R^2 = 0.87$, demonstrating that a single electrostatic quantity can
unify three independent chemical concepts. However, like all previous
scales, it remains static: it describes the neutral atom in its
ground-state configuration and provides no information about how the
electronegativity changes when the atom participates in bonding.

Here we remove this limitation by extending the AMIP to fractional
charge states. We promote the mean inner potential to a function
$\chi(\delta)$ of the fractional positive charge $\delta$, and
expand it in a Taylor series around the neutral state. We work
throughout in the square-root representation
\(\chi\equiv\chi_{\mathrm{AMIP},1/2}\), which is the primary scale of
the static AMIP framework and carries the closest correspondence to
the classical electronegativity scales; the linear (\(p=1\))
representation is retained in the Supplemental Material and is related
to the square-root form by the exact transformation
\(\chi^{(1/2)}=\sqrt{\chi^{(1)}}\). The zeroth-order term recovers the
static scale; the first-order term defines the
\textbf{Electrostatic Rigidity} \(\mathcal{R}^{(1/2)}\), whose
\textit{sign} directly indicates whether the atom hardens or softens
upon oxidation; and the second-order term defines the
\textbf{Electrostatic Anharmonic Modulus} \(\mathcal{A}^{(1/2)}\),
which governs whether the response accelerates or saturates.

A logarithmic-derivative decomposition reveals that the macroscopic
response coefficients factorize into a static scale and a
scale-independent, dimensionless quantity:
\begin{equation}
\mathcal{R}^{(p)} = c_p \, \chi_0^{(p)} \, \mathcal{D},
\qquad
\mathcal{D} \equiv \frac{\mathcal{R}_\Omega}{\Omega_0}
- \frac{\mathcal{R}_f}{f_0},
\label{eq:D_intro}
\end{equation}
where \(c_{1}=1\) and \(c_{1/2}=1/2\). We call \(\mathcal{D}\) the
\textbf{intrinsic response asymmetry}: it is independent of the scale
exponent \(p\), and its sign and magnitude determine whether an atom
softens or hardens and how strongly. The static scale
\(\chi_0^{(p)}\) sets the overall amplitude, while \(\mathcal{D}\)
carries the new physics of the response. The two terms in
\(\mathcal{D}\) correspond to two competing microscopic
channels---electronic screening loss, which favors softening, and
geometric volume collapse, which favors hardening. Applying this
framework to all 102 elements, we find that this competition produces
a systematic dichotomy between softening main-group elements and
hardening d-block elements, two previously unrecognized scaling laws
within the transition series, and two qualitatively distinct
anomalies: a seesaw at half-filled d\(^5\) configurations driven by
the sensitivity of the exchange-stabilized d\(^5\) shell to changes in
its \(s\)-electron screening layer, and a near-perfect cancellation of
electronic and geometric responses in Ce. The framework also resolves
the century-old Cs--Fr controversy through a systematic
non-relativistic, scalar-relativistic, and full-relativistic
comparison, revealing Fr as a \textit{critical element} whose
near-boundary response is amplified fivefold by relativistic
contraction of the 7s orbital.

This work thus transforms electronegativity from a lookup-table number
into a continuous, hierarchically structured, and experimentally
falsifiable quantum descriptor. It bridges three communities that have
developed largely independently---condensed-matter physics (through the
mean inner potential and collective charge response), quantum chemistry
(through conceptual DFT and chemical hardness), and materials science
(through the design of catalysts and redox-stable compounds). The
remainder of this paper is organized as follows.
Section~\ref{sec:theory} develops the theoretical framework, the
logarithmic-derivative decomposition, and the intrinsic response
asymmetry \(\mathcal{D}\).
Section~\ref{sec:methods} describes the computational methodology.
Section~\ref{sec:results} presents the periodic-table results, the
three scaling laws, the two anomalies, and the Cs/Fr resolution.
Section~\ref{sec:conclusions} summarizes the key findings and outlines
future directions.

\section{Theoretical Framework}
\label{sec:theory}

\subsection{The fractional-charge AMIP functional}
\label{sec:theory_functional}

The static AMIP electronegativity is defined as~\cite{Zheng2026AMIP}
\begin{equation}
\chi_{\mathrm{AMIP},p} = C_p \left( \frac{v_0}{n_q} \right)^{p},
\label{eq:chi0_general}
\end{equation}
where $p$ is a positive exponent, $n_q$ is the principal quantum
number of the valence shell, $C_p$ is a normalization constant, and
$v_0 = K f_e(0) / \Omega_a$ is the atomic mean inner potential. Here
$f_e(0)$ is the forward electron scattering factor, and $\Omega_a$ the
atomic volume. These quantities are computed from the total charge
density $\rho_t(\mathbf{r})$ and the valence charge density
$\rho_v(\mathbf{r})$, respectively. Following Ref.~\cite{Zheng2026AMIP},
we adopt the mean valence radius
\begin{equation}
r_v = \frac{\int r\, \rho_v(\mathbf{r}) \, d^3 r}{\int \rho_v(\mathbf{r}) \, d^3 r}
\label{eq:rv}
\end{equation}
and set $\Omega_a = 4\pi r_v^3/3$. The constant
$K = h^2/(2\pi m_0 e) = 47.87658$ eV$\cdot$\AA$^2$ ensures that $v_0$
is expressed in volts~\cite{Peng1999}. Throughout this work we use
the $p = 1/2$ (square-root) scale as the primary representation, which
is consistent with the static AMIP framework of
Ref.~\cite{Zheng2026AMIP} and carries the closest correspondence to
the classical electronegativity scales. The $p = 1$ (linear) scale,
which carries direct physical units (V) and maps transparently onto the
chemical potential and hardness of conceptual DFT, is retained in the
Supplemental Material for comparison; all qualitative conclusions are
independent of \(p\), as made explicit by the factorization in
Eq.~\eqref{eq:D_intro} above.

To extend this framework to fractional charge states, we promote
$v_0$ to a function of the fractional positive charge $\delta$,
defined as the number of electrons removed from the neutral atom
($N_e = Z - \delta$):
\begin{equation}
v_0(\delta) = K \frac{f_e(0; \delta)}{\Omega(\delta)},
\label{eq:v0_delta}
\end{equation}
where
\begin{equation}
f_e(0; \delta) = \frac{1}{3} \int_0^\infty \, r^2 \,\rho_{t,\delta}(\mathbf{r})  d^3r
\label{eq:fe_full}
\end{equation}
is the forward scattering factor in \AA~\cite{Peng1999,Zheng2009}, and
$\Omega_a(\delta) = 4\pi r_v(\delta)^3/3$ is obtained from the mean
valence radius
\begin{equation}
r_v(\delta) = \frac{\int_0^\infty r\, \rho_{v,\delta}(\mathbf{r}) \, d^3r}{\int_0^\infty \rho_{v,\delta}(\mathbf{r}) \, d^3r}.
\label{eq:rv_def}
\end{equation}
All integrals run over the full space, so no empirical cutoff enters
the definition. The fractional-occupancy density
$\rho_{\delta}(\mathbf{r})$ is computed self-consistently with the
all-electron atomic DFT code \texttt{ld1.x}~\cite{QE}, using the
full relativistic (four-component) treatment for all elements.

\subsection{Hierarchical expansion and the intrinsic response
asymmetry}
\label{sec:theory_decomposition}

Expanding \(\chi(\delta) \equiv \chi_{\mathrm{AMIP},1/2}(\delta)\) in
a Taylor series around the neutral state gives
\begin{equation}
\chi(\delta) = \chi_0 + \mathcal{R} \delta + \frac{1}{2} \mathcal{A} \delta^2 + \mathcal{O}(\delta^3),
\label{eq:taylor}
\end{equation}
where, for notational economy, we henceforth write
\(\chi_0\equiv\chi_{\mathrm{AMIP},1/2}(0)\),
\(\mathcal{R}\equiv\mathcal{R}^{(1/2)}\), and
\(\mathcal{A}\equiv\mathcal{A}^{(1/2)}\) unless explicitly stated
otherwise. The zeroth-order term is the static AMIP
electronegativity---the potential-level analog of the chemical
potential $\mu$. The first-order term
\begin{equation}
\mathcal{R} = \left. \frac{d \chi}{d \delta} \right|_{\delta=0}
\label{eq:R_def}
\end{equation}
is the \textbf{Electrostatic Rigidity}. Unlike any static scale,
$\mathcal{R}$ carries a \textit{sign}: $\mathcal{R} > 0$ signals
ionization-induced hardening (the potential deepens upon electron
removal), whereas $\mathcal{R} < 0$ signals softening. It is the
potential-level analog of chemical hardness $\eta$. The second-order
term
\begin{equation}
\mathcal{A} = \left. \frac{d^2 \chi}{d \delta^2} \right|_{\delta=0}
\label{eq:A_def}
\end{equation}
is the \textbf{Electrostatic Anharmonic Modulus}, which determines
whether the response accelerates ($\mathcal{A} > 0$) or saturates
($\mathcal{A} < 0$) as charge is progressively removed. It is the
potential-level analog of hyper-hardness $\gamma$.

The physical content of $\mathcal{R}$ and $\mathcal{A}$ is best
revealed by a logarithmic-derivative decomposition~\cite{zheng2006}.
Because $n_q$ is independent of $\delta$, it drops out of the
logarithmic derivative and does not enter the response coefficients.
Because $\chi(\delta)\propto
[f_e(0;\delta)/\Omega_a(\delta)]^{1/2}$ in the \(p=1/2\) scale, the
relative change of $\chi$ is one half the difference between the
relative changes of its two microscopic ingredients:
\begin{equation}
\frac{d}{d\delta} \ln \chi(\delta)
= \frac{1}{2}\left(
\frac{f'_e(0;\delta)}{f_e(0;\delta)}
- \frac{\Omega_a'(\delta)}{\Omega_a(\delta)}
\right).
\label{eq:logderiv}
\end{equation}
Evaluating at $\delta = 0$ and introducing the shorthand
$f_0 = f_e(0;0)$, $f_1 = f'_e(0)$, $f_2 = f''_e(0)$,
$\Omega_0 = \Omega_a(0)$, $\Omega_1 = \Omega_a'(0)$,
$\Omega_2 = \Omega_a''(0)$, we obtain
\begin{equation}
\mathcal{R} = \frac{1}{2}\,\chi_0
\left( \frac{f_1}{f_0} - \frac{\Omega_1}{\Omega_0} \right).
\label{eq:R_raw}
\end{equation}

Following the materials-science convention in which stiffness
coefficients are defined with a minus sign so that they are positive
for a stable solid ($K = -V\,dP/dV$), we define the electronic and
geometric rigidity coefficients
\begin{equation}
\mathcal{R}_f = -f_1, \qquad \mathcal{R}_\Omega = -\Omega_1,
\label{eq:RfRomega}
\end{equation}
and introduce the \textbf{intrinsic response asymmetry}
\begin{equation}
\boxed{\;
\mathcal{D} \equiv \frac{\mathcal{R}_\Omega}{\Omega_0}
- \frac{\mathcal{R}_f}{f_0},
\qquad
\mathcal{R} = \frac{1}{2}\, \chi_0 \, \mathcal{D}.
\;}
\label{eq:D_def}
\end{equation}
The two terms have transparent physical meaning: the geometric term
$+\mathcal{R}_\Omega/\Omega_0$ represents the hardening contribution
from volume collapse, while the electronic term
$-\mathcal{R}_f/f_0$ represents the softening contribution from
screening loss. The sign of \(\mathcal{D}\) is therefore determined by
the competition between these two channels; its magnitude measures
how strongly the winning channel dominates. Crucially, \(\mathcal{D}\)
is a purely microscopic quantity constructed from the fitted
coefficients of \(f_e(0;\delta)\) and \(\Omega(\delta)\), and is
therefore \textit{independent of the scale exponent \(p\)}.
Under the normalization $\chi_{\mathrm{AMIP},1}=
(\chi_{\mathrm{AMIP},1/2})^2$ adopted in Ref.~\cite{Zheng2026AMIP},
the macroscopic response coefficients in the two representations are
related by
\begin{equation}
\mathcal{R}^{(1/2)} = \frac{1}{2}\, \chi_0^{(1/2)} \, \mathcal{D},
\qquad
\mathcal{R}^{(1)} = \chi_0^{(1)} \, \mathcal{D},
\label{eq:R_p_relation}
\end{equation}
so that all qualitative statements about the sign and relative
strength of the response are statements about \(\mathcal{D}\), while
the static scale sets the overall amplitude. This is the central
structural feature of the present framework: \(\mathcal{D}\) isolates
the new physics of the response from the static information already
encoded in \(\chi_0^{(p)}\).

The same procedure applied to the second derivative yields the
corresponding decomposition of the anharmonic modulus in the square-root
representation:
\begin{equation}
\boxed{\;
\mathcal{A} = \chi_0 \Bigg[
\frac{1}{2} \left( \frac{\mathcal{A}_\Omega}{\Omega_0}
- \frac{\mathcal{A}_f}{f_0} \right)
+ \frac{1}{2} \left( \frac{\mathcal{R}_\Omega^2}{\Omega_0^2}
- \frac{\mathcal{R}_f^2}{f_0^2} \right)
+ \frac{1}{4} \mathcal{D}^2 \Bigg], \;}
\label{eq:A_decomp}
\end{equation}
with \(\mathcal{A}_f = -f_2\) and \(\mathcal{A}_\Omega = -\Omega_2\).
The analogous expression for the \(p = 1\) scale,
written in the same structural form to make the correspondence
explicit, is
\begin{equation}
\mathcal{A}^{(1)} = \chi_0^{(1)} \left[
\left( \frac{\mathcal{A}_\Omega}{\Omega_0}
- \frac{\mathcal{A}_f}{f_0} \right)
+ \left( \frac{\mathcal{R}_\Omega^2}{\Omega_0^2}
- \frac{\mathcal{R}_f^2}{f_0^2} \right)
+ \mathcal{D}^2 \right].
\label{eq:A_decomp_p1}
\end{equation}
Comparing Eqs.~\eqref{eq:A_decomp} and~\eqref{eq:A_decomp_p1}, the
three terms in the square-root representation are exactly one half,
one half, and one quarter of their linear-scale counterparts,
reflecting the suppression of the second-order response by the
square-root exponent.

Equations~\eqref{eq:R_raw} and~\eqref{eq:A_decomp} are used as the
operational definitions of $\mathcal{R}$ and $\mathcal{A}$: the
macroscopic response coefficients are obtained analytically from the
fitted coefficients of $f_e(0;\delta)$ and $\Omega(\delta)$, rather
than from a separate fit of $\chi(\delta)$ itself. The Taylor
expansion of Eq.~\eqref{eq:taylor} then provides their interpretation
as the first and second derivatives of $\chi$ at $\delta=0$.

The full set of \(\mathcal{R}\), \(\mathcal{R}^{(1)}\),
\(\mathcal{A}\), and \(\mathcal{A}^{(1)}\) values for all 102 elements,
together with their electronic and geometric decompositions and the
intrinsic response asymmetry \(\mathcal{D}\), is tabulated in
Tables S1, S2, and S5 in the Supplemental Material. The fitting
results for $f_e(0)$ and $\Omega_a$ are listed in Tables S3 and S4,
respectively.

Finally, we note that the Taylor expansion can be extended to higher
orders, $\chi(\delta) = \chi_0 + \mathcal{R}\delta +
\frac{1}{2}\mathcal{A}\delta^2 + \frac{1}{6}\mathcal{B}\delta^3 +
\frac{1}{24}\mathcal{C}\delta^4 + \cdots$, where $\mathcal{B}$ and
$\mathcal{C}$ correspond to the hyper-hardness and higher-order
hyper-hardnesses of conceptual DFT~\cite{parr1989density}. We restrict
the present analysis to second order, which captures the dominant
physics for the chemical window $0 \le \delta \le 0.3$; higher-order
terms become relevant only for strongly correlated systems, and are
discussed as a future direction in Sec.~\ref{sec:conclusions}.

\subsection{Variational interpretation of the response hierarchy}
\label{sec:theory_variational}

The response coefficients admit a variational interpretation that
makes transparent why the logarithmic-derivative decomposition of
Eq.~\eqref{eq:logderiv} is the natural language for the hierarchy.
The AMIP is itself a composite functional of the electron density,
built from two elementary density functionals. The first is the
forward scattering factor
\begin{equation}
F_1[\rho] \equiv \frac{1}{3}\int_0^\infty r^2 \rho_t(\mathbf{r})\, d^3r,
\label{eq:F1}
\end{equation}
which is linear in the total density $\rho_t$. The second is the mean
valence radius
\begin{equation}
r_v[\rho] \equiv
\frac{\int_0^\infty r\, \rho_v(\mathbf{r})\, d^3r}
{\int_0^\infty \rho_v(\mathbf{r})\, d^3r},
\qquad
F_2[\rho] \equiv \frac{4\pi}{3}\bigl(r_v[\rho]\bigr)^3,
\label{eq:F2}
\end{equation}
which depends only on the valence density $\rho_v$ and is
homogeneous of degree zero in its normalization. The AMIP is then
the ratio
\begin{equation}
v_0[\rho] = K \frac{F_1[\rho]}{F_2[\rho]},
\label{eq:v0_functional}
\end{equation}
and the AMIP electronegativity is the composite functional
\begin{equation}
\chi[\rho] = \frac{C_{1/2}}{n_q^{1/2}}\, v_0[\rho]^{1/2}
= \left(\frac{C_{1/2}^2 K}{n_q}\right)^{1/2}\,
\left(\frac{F_1[\rho]}{F_2[\rho]}\right)^{1/2}.
\label{eq:chi_functional}
\end{equation}
For a fixed fractional charge $\delta$, the equilibrium density
$\rho_\delta$ is determined by the standard density-functional
variational principle under the constraint
$\int \rho\, d^3r = Z - \delta$~\cite{hohenberg1964inhomogeneous,kohn1965self,parr1989density}. Substituting $\rho_\delta$ into
Eq.~\eqref{eq:chi_functional} defines the curve $\chi(\delta)$ whose
Taylor coefficients are $\mathcal{R}$ and $\mathcal{A}$.

Because $\chi[\rho]$ is a power of a ratio of two density functionals,
its logarithm separates the two elementary channels:
\begin{equation}
2\ln \chi[\rho] = \ln\!\left(\frac{C_{1/2}^2 K}{n_q}\right)
+ \ln F_1[\rho] - \ln F_2[\rho].
\label{eq:log_chi_functional}
\end{equation}
Differentiating along the constrained equilibrium path
$\rho_\delta$ gives
\begin{equation}
2\frac{d}{d\delta}\ln \chi[\rho_\delta]
= \frac{1}{F_1}\frac{d F_1}{d\delta}
- \frac{1}{F_2}\frac{d F_2}{d\delta},
\label{eq:log_chi_path}
\end{equation}
which is exactly the logarithmic-derivative decomposition of
Eq.~\eqref{eq:logderiv}, now expressed as the difference between the
relative responses of the two elementary functionals. Evaluating at
$\delta = 0$, and using $F_1[\rho_0]=f_0$ and $F_2[\rho_0]=\Omega_0$,
gives the microscopic expression for $\mathcal{D}$ entering
Eq.~\eqref{eq:D_def}:
\begin{equation}
\mathcal{D} = \frac{\mathcal{R}_\Omega}{\Omega_0}
- \frac{\mathcal{R}_f}{f_0},
\qquad
\mathcal{R}_f = -\frac{d F_1}{d\delta}\bigg|_0,
\quad
\mathcal{R}_\Omega = -\frac{d F_2}{d\delta}\bigg|_0.
\label{eq:D_from_functionals}
\end{equation}
The second derivative along the same path yields
Eq.~\eqref{eq:A_decomp}, with
$\mathcal{A}_f = -d^2F_1/d\delta^2|_0$ and
$\mathcal{A}_\Omega = -d^2F_2/d\delta^2|_0$.

The response hierarchy $\chi_0 \to \mathcal{R} \to \mathcal{A}$ is
therefore not an arbitrary parametrization but the Taylor expansion
of a composite density functional along the constrained
fractional-charge coordinate. The electronic channel
$\mathcal{R}_f/f_0$ is the relative response of the linear functional
$F_1[\rho]$, which measures how the electron cloud contracts in
real space; the geometric channel
$\mathcal{R}_\Omega/\Omega_0$ is the relative response of the
nonlinear functional $F_2[\rho]$, which measures how the valence
radius contracts. The competition between these two elementary
responses, encoded in the single scale-independent quantity
\(\mathcal{D}\), is the variational origin of the
softening--hardening dichotomy.

\subsection{Relationship to conceptual DFT}
\label{sec:theory_cdft}

The hierarchical response theory is complementary to conceptual
DFT~\cite{ParrDonnellyLevyPalke1978,ParrPearson1983,Pearson1963}.
Table~\ref{tab:CDFT_mapping} summarizes the formal order-by-order
correspondence between the two frameworks. The central difference is
the field quantity being differentiated: CDFT descriptors are
derivatives of the \textit{total energy} $E(N)$, whereas the AMIP
coefficients are derivatives of the \textit{electronegativity}
$\chi(N)\propto v_0(N)^{1/2}$. The former describe thermodynamic
driving forces and energy costs; the latter describe real-space
electrostatic stiffness. Crucially, while the energy derivatives are
difficult to isolate experimentally, the AMIP coefficients map
directly onto the Coulomb potential accessible through quantitative
electron holography~\cite{Mkhoyan2008,Gajdardziska1993,Schowalter2002},
making them experimentally falsifiable at the atomic scale. We
emphasize that this correspondence is conceptual rather than
quantitative; establishing a numerical mapping between the AMIP
coefficients and the CDFT descriptors is an interesting direction for
future work but is not required for the present framework.

The correspondence in Table~\ref{tab:CDFT_mapping} is formal: it
matches the order of the Taylor expansion. Physically, however, the
mapping is shifted by one derivative. Because the AMIP is an
electrostatic potential rather than a total energy, the static AMIP
electronegativity $\chi_0$ already plays the role of the chemical
potential $\mu$ (or its negative), not that of the total energy
$E(N)$. Correspondingly, the Electrostatic Rigidity $\mathcal{R}$
plays the role of chemical hardness $\eta$, and the Anharmonic
Modulus $\mathcal{A}$ plays the role of hyper-hardness $\gamma$. This
one-derivative shift reflects the fact that the AMIP is a
potential-level, rather than energy-level, descriptor. Both mappings
are useful: the formal one organizes the hierarchy of coefficients,
while the physical one clarifies their chemical meaning. The
intrinsic response asymmetry \(\mathcal{D}\) has no direct CDFT
analog; it is the scale-independent ratio that separates the new
response information from the static information already contained in
\(\chi_0\).

\begin{table}[H]
\centering
\footnotesize
\renewcommand{\tabcolsep}{4pt}
\caption{Formal order-by-order correspondence between conceptual DFT
(energy derivatives with respect to electron number) and the
hierarchical response theory (electronegativity derivatives with
respect to fractional charge $\delta=Z-N$). The table lists the
\emph{formal} correspondence only; the \emph{physical} mapping is
shifted by one derivative, with $\chi_0\leftrightarrow\mu$,
$\mathcal{R}\leftrightarrow\eta$, and
$\mathcal{A}\leftrightarrow\gamma$ (see text). The third-order
coefficient $\mathcal{B}$ is defined in Sec.~II B but is not used in
the present analysis.}
\label{tab:CDFT_mapping}
\begin{tabular}{c l l}
\toprule
\textbf{Order} & \textbf{Conceptual DFT} & \textbf{This Work} \\
\midrule
0 & $E(N)$ (total energy) & $\chi_0 = \chi_{\mathrm{AMIP},1/2}(0)$ \\
1 & $\mu = \partial E/\partial N$ (chemical potential) & $\mathcal{R} = d\chi/d\delta$ (electrostatic rigidity) \\
2 & $\eta = \frac{1}{2}\partial^2 E/\partial N^2$ (hardness) & $\mathcal{A} = d^2\chi/d\delta^2$ (anharmonic modulus) \\
3 & $\gamma = \partial^3 E/\partial N^3$ (hyper-hardness) & $\mathcal{B} = d^3\chi/d\delta^3$ \\
\bottomrule
\end{tabular}
\end{table}

\section{Computational Methodology}
\label{sec:methods}

All calculations were performed with the all-electron atomic DFT code
\texttt{ld1.x} from the Quantum ESPRESSO distribution~\cite{QE},
which solves the radial Kohn--Sham equation for the spherically
symmetric atomic potential. We used the revPBE generalized gradient
approximation for exchange and correlation, following our previous
static AMIP work~\cite{Zheng2026AMIP}, and the full relativistic
(four-component Dirac) treatment for all elements, which includes both
scalar relativistic effects and spin-orbit coupling. Fractional charge
states were generated by specifying the fractional occupation number
$f_i$ of each valence orbital, subject to the constraint
$\sum_i f_i = Z_v - \delta$, where $Z_v$ is the number of valence
electrons in the neutral atom. For each value of $\delta$, the
self-consistent field equations were converged to $10^{-15}$ Ry,
and the resulting valence density $\rho_\delta(\mathbf{r})$ was used
to compute the two independent density-derived quantities that enter
the AMIP functional, the forward scattering factor $f_e(0;\delta)$
and the mean valence radius $r_v(\delta)$, through
Eqs.~\eqref{eq:fe_full} and~\eqref{eq:rv_def}; the atomic volume
follows as $\Omega_a(\delta) = 4\pi r_v(\delta)^3/3$. The mean inner
potential was then obtained as
$v_0(\delta) = K f_e(0;\delta)/\Omega_a(\delta)$ with
$K = 47.87658$ eV$\cdot$\AA$^2$. For each element we computed the
seven fractional charge states $\delta = 0, 0.05, 0.1, 0.15, 0.2,
0.25, 0.3$.
Electronic configurations are the reference configurations of
Ref.~\cite{Zheng2026AMIP}; the few documented exceptions (e.g., the
\(d^{10}s^{1}\) configuration for Cu, Ag, and Au, and the
\(4d^{9}5s^{1}\) configuration for Pd) are listed in the caption of
Table~S1.

The response coefficients were obtained analytically from the fitted
coefficients of the two microscopic ingredients. Specifically,
$f_e(0;\delta)$ and $\Omega(\delta)$ were each fitted over the
oxidative window $0 \le \delta \le 0.3$ with the quadratic form
$Y(\delta) = y_0 + y_1\delta + \frac{1}{2}y_2\delta^2$, yielding
the coefficients $\{f_0, f_1, f_2\}$ and
$\{\Omega_0, \Omega_1, \Omega_2\}$. Substituting these into
Eqs.~\eqref{eq:D_def} and~\eqref{eq:A_decomp} gives the intrinsic
response asymmetry \(\mathcal{D}\) and the response coefficients
$\mathcal{R}$ and $\mathcal{A}$ directly, without any further
fitting; the \(p=1\) coefficients follow from the second relation in
Eq.~\eqref{eq:R_p_relation}. Because $f_e(0;\delta)$ and
$\Omega(\delta)$ each carry a strict physical meaning---the forward
scattering factor and the atomic volume, respectively---this
analytical route ensures that the decomposition into electronic and
geometric contributions is numerically consistent and physically
transparent. The resulting coefficients for all 102 elements are
tabulated in Tables~S1 and~S2 for the \(p=1/2\) and \(p=1\) scales,
respectively, and the decomposition into electronic and geometric
channels is given in Table~S5. 
The quadratic form is checked directly against the seven computed
$\chi(\delta)$ points for every element; it reproduces them to high
accuracy over the oxidative window, as expected from the smoothness of
the underlying density response.

The fitting window is restricted to the oxidative branch
($\delta \ge 0$) for two reasons. First, this maintains a uniform
valence-shell reference state across the entire periodic table: for
open-shell atoms, electron removal occurs within the same valence
shell, whereas for closed-shell atoms, the oxidative branch avoids
the orbital-shell discontinuity that would arise from electron
addition. Second, extensive DFT studies of charge transfer in
molecular and crystalline environments indicate that the net charge
transferred to or from a bonded atom rarely exceeds $0.3\,e$
\cite{bader1990atoms,marenich2012charge}, so the oxidative window
$0 \le \delta \le 0.3$ covers the chemically relevant range.


\section{Results and Discussion}
\label{sec:results}

\subsection{The electrostatic response spectrum and the
softening--hardening dichotomy}
\label{sec:results_spectrum}

\begin{figure}[htbp]
\centering
\includegraphics[width=0.9\linewidth]{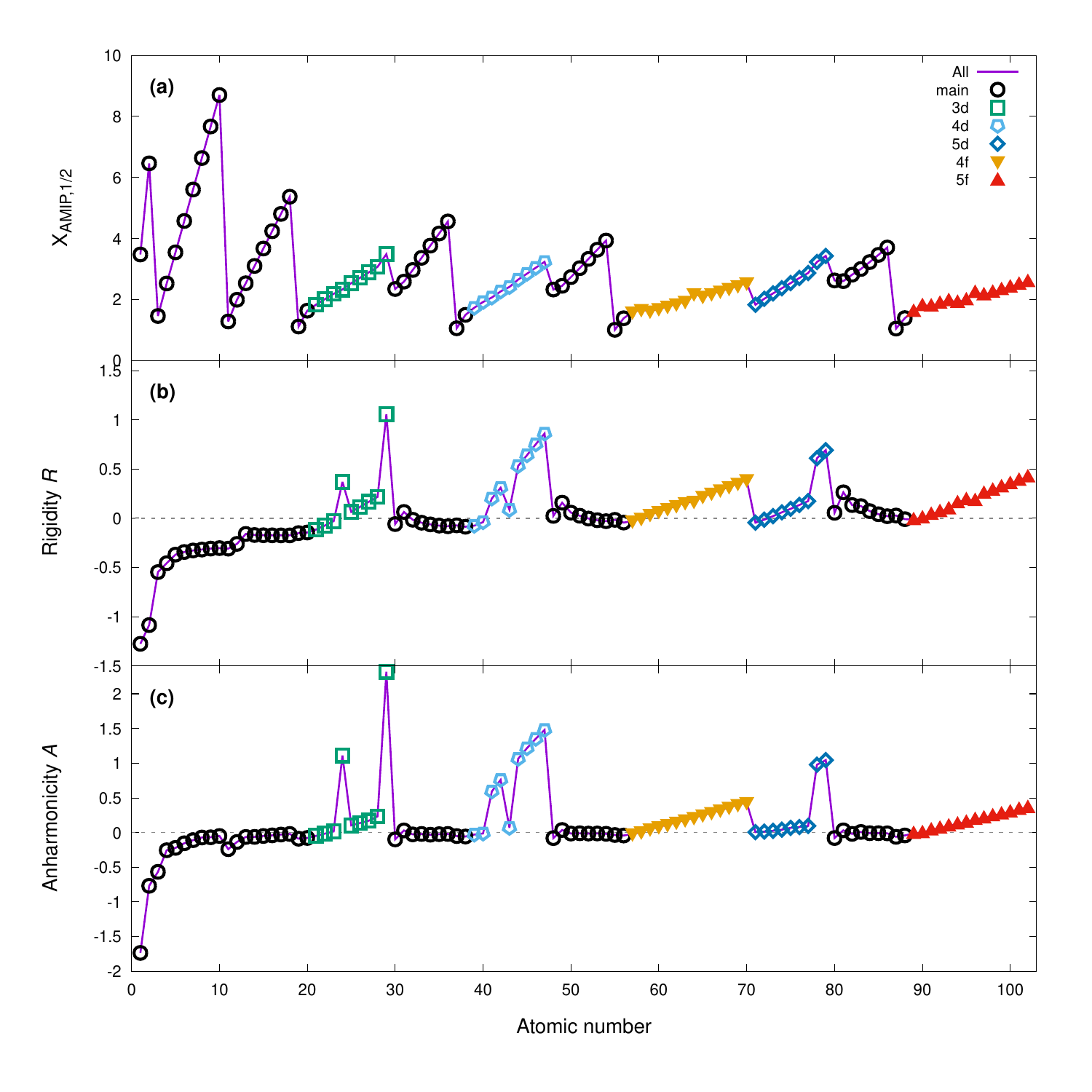}
\caption{Atomic-level electronic descriptors for the \(p=1/2\) scale
derived from the atomic mean inner potential (AMIP). In all panels,
main-group elements are denoted by open circles, \(3d\) by open
squares, \(4d\) by open pentagons, \(5d\) by open diamonds, \(4f\) by
solid inverted triangles, and \(5f\) by solid upright triangles.
\textbf{(a)}~AMIP-based electronegativity
\(\chi_{\mathrm{AMIP},1/2}\) plotted as a function of atomic number.
The curve is identical to the established AMIP scale reported in
Ref.~\cite{Zheng2026AMIP}.
\textbf{(b)}~Electrostatic rigidity \(\mathcal{R}\). Positive
\(\mathcal{R}\) indicates ionization-induced hardening; negative
\(\mathcal{R}\) indicates softening.
\textbf{(c)}~Electrostatic anharmonicity \(\mathcal{A}\). Positive
\(\mathcal{A}\) indicates accelerating response; negative
\(\mathcal{A}\) indicates saturation.}
\label{fig:amip_descriptors}
\end{figure}

Calculating the response coefficients for all 102 elements reveals a
clear picture of how atomic electronegativity responds to charge
perturbation. Figure~\ref{fig:amip_descriptors} presents the three
descriptors---the static electronegativity
$\chi_0=\chi_{\mathrm{AMIP},1/2}$, the electrostatic rigidity
$\mathcal{R}$, and the anharmonic modulus $\mathcal{A}$---as
functions of atomic number. The static scale in panel~(a) reproduces
the established AMIP result~\cite{Zheng2026AMIP}, with maxima at noble
gases and minima at alkali metals. The response coefficients, however,
reveal a structure that no static scale can capture. Panel~(b) shows
that $\mathcal{R}$ spans the full range from strong softening
($\mathcal{R} = -1.09$ V$^{1/2}$/$e$ for He) to strong hardening
($\mathcal{R} = +1.09$ V$^{1/2}$/$e$ for Cu), and that its sign
divides the periodic table into two broadly distinct regimes:
main-group elements predominantly soften, while d-block elements
harden. Panel~(c) shows that $\mathcal{A}$ follows a similar pattern
but with a different distribution: the closed-shell d$^{10}$
configurations of the late transition metals, including the coinage
metals, exhibit the largest positive values, reflecting the absence
of low-energy electron-reorganization channels. The noble gases, by
contrast, exhibit only small negative values, because their valence
shells are already tightly bound and offer no low-energy channel for
further contraction.

\begin{figure*}[t]
\centering
\includegraphics[width=\textwidth]{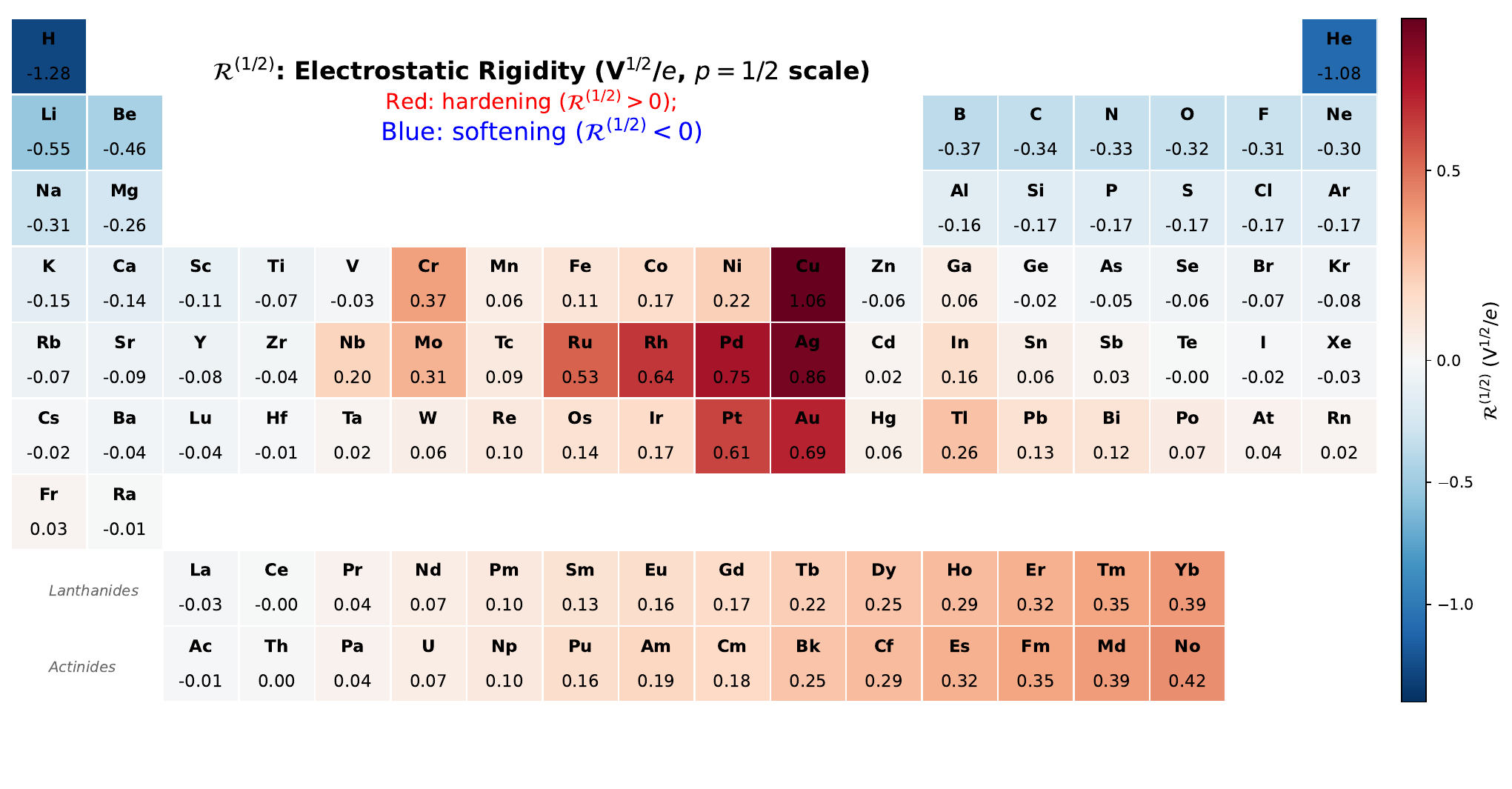}
\caption{Periodic-table organization of the intrinsic response
asymmetry \(\mathcal{D}\) (dimensionless) and, for orientation, the
electrostatic rigidity
\(\mathcal{R}^{(1/2)}\) (V$^{1/2}$/$e$). Red indicates hardening
(\(\mathcal{D}>0\)); blue indicates softening (\(\mathcal{D}<0\));
pale cells correspond to near-zero response. The block-level
structure---uniform softening of the \(s\)- and \(p\)-blocks,
systematic hardening of the \(d\)-block, and the transitional
behavior of the \(f\)-block---is directly visible. The near-zero
values at Cs, Fr, and Ce, discussed in
Secs.~\ref{sec:results_anomalies} and~\ref{sec:results_csfr},
appear as pale cells. Because \(\mathcal{D}\) is independent of the
scale exponent \(p\), this map is identical in structure to the
\(p=1\) map shown in Fig.~S9 of the Supplemental Material.}
\label{fig:D_map}
\end{figure*}

The intrinsic response asymmetry \(\mathcal{D}\) makes the same
structure visible in a scale-independent way. Figure~\ref{fig:D_map}
displays \(\mathcal{D}\) across the periodic table, and
Table~\ref{tab:decomp} lists the decomposition for a representative
set of elements. \(\mathcal{D}\) ranges from \(-0.734\) (H) to
\(+0.626\) (Cu), and its sign divides the elements exactly as
\(\mathcal{R}\) does, because \(\chi_0>0\) and \(c_{1/2}>0\). The
advantage of \(\mathcal{D}\) is that it isolates the competition
between the electronic and geometric channels from the amplitude set
by the static scale. For example, Li
(\(\mathcal{D}=-0.749\), \(\mathcal{R}=-0.548\) V$^{1/2}$/$e$) and
Cs (\(\mathcal{D}=-0.031\), \(\mathcal{R}=-0.016\)
V$^{1/2}$/$e$) are both softening, but the comparison of their
\(\mathcal{D}\) values shows that Li's response is intrinsically far
more asymmetric than Cs's, whereas a comparison of \(\mathcal{R}\)
values alone would conflate this intrinsic difference with the
difference in static scale. Similarly, Fr
(\(\mathcal{D}=+0.051\), \(\mathcal{R}=+0.027\)
V$^{1/2}$/$e$) and Ce (\(\mathcal{D}=-0.001\),
\(\mathcal{R}=-0.001\) V$^{1/2}$/$e$) both lie near the boundary,
but \(\mathcal{D}\) shows that Fr is decisively on the hardening side
while Ce is almost exactly critical.

Hydrogen and helium define the extreme softening boundary of the
periodic table, with \(\mathcal{R} = -1.277\) and \(-1.090\)
V$^{1/2}$/$e$, respectively (\(\mathcal{D} = -0.734\) and
\(-0.337\)). Both elements lack inner-shell electrons to buffer the
radial expansion of the remaining electron cloud, resulting in a
precipitous drop in \(\chi(\delta)\) at small \(\delta\) followed by
rapid saturation, as evidenced by their large negative \(\mathcal{A}\)
values. No element with inner-shell electrons approaches this regime,
which highlights a fundamental limitation of static scales: by
assigning a single number to each element, they cannot distinguish the
extreme softening of H and He from the moderate softening of the
heavier main-group elements.

\begin{figure}[htbp]
\centering
\includegraphics[width=\linewidth]{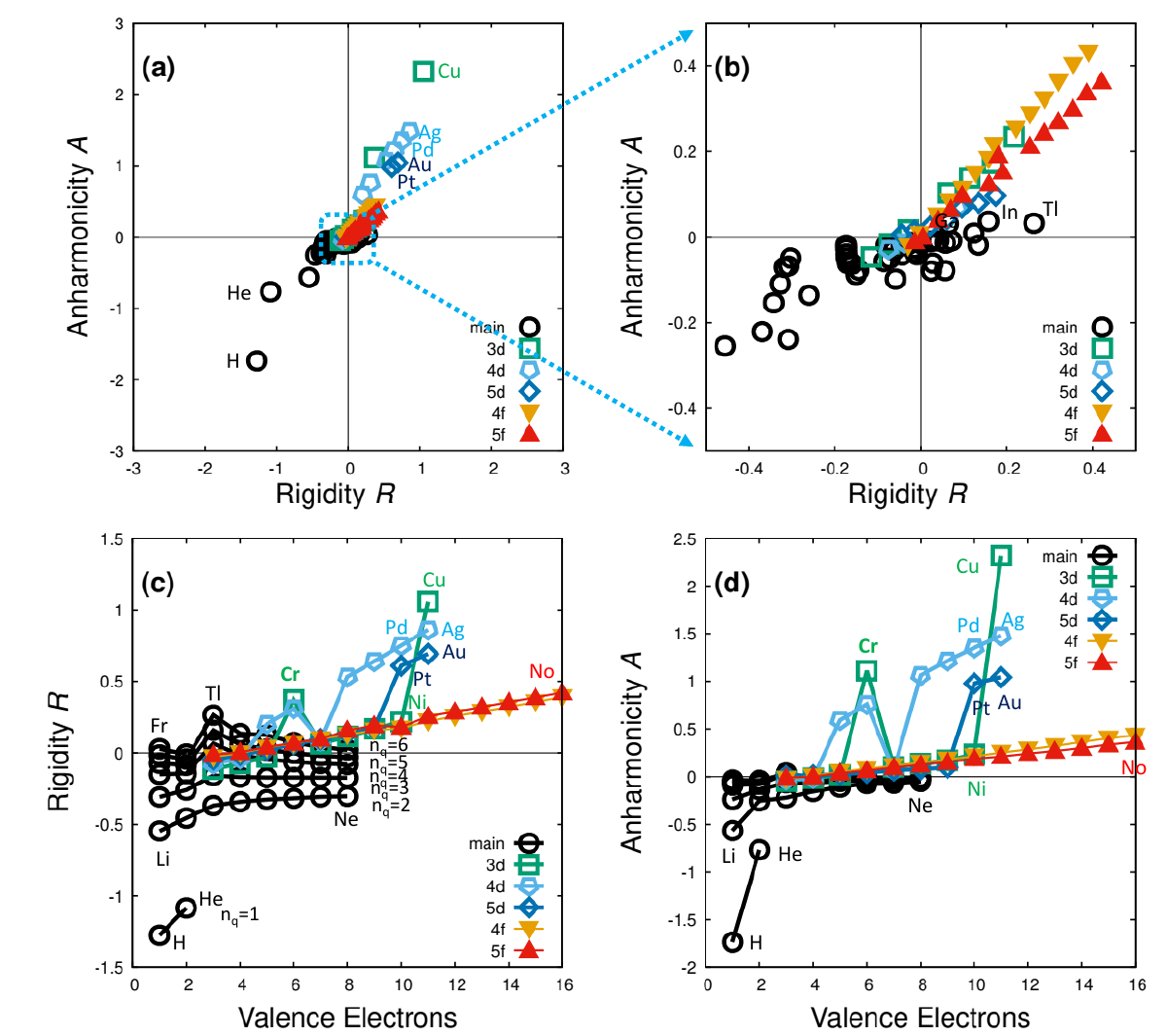}
\caption{Response space and periodic-table organization of the AMIP
response coefficients for the \(p=1/2\) scale.
\textbf{(a)}~Parametric correlation of \(\mathcal{A}\) versus
\(\mathcal{R}\) for all 102 elements.
\textbf{(b)}~Magnified view of the near-origin region.
\textbf{(c)}~\(\mathcal{R}\) plotted against the number of valence
electrons.
\textbf{(d)}~\(\mathcal{A}\) plotted against the number of valence
electrons.}
\label{fig:response_space}
\end{figure}

Figure~\ref{fig:response_space} reorganizes the same data to expose
the structure of the response space. Panel~(a) plots \(\mathcal{A}\)
against \(\mathcal{R}\) and reveals that the 102 elements are not
randomly distributed: main-group elements cluster in the lower-left
quadrant (softening with saturation), while d-block elements occupy
the upper-right quadrant (hardening with acceleration). The f-block
elements span the diagonal, bridging the two regions. The near-origin
magnification in panel~(b) resolves the boundary cases: Fr
(\(\mathcal{D} \approx +0.051\)) and Cs (\(\mathcal{D} \approx
-0.031\)) lie at opposite edges of the main-group cluster, while Te
(\(\mathcal{D} \approx -0.002\)) and Ce (\(\mathcal{D} \approx
-0.001\)) appear as near-zero outliers in their respective blocks.
Panels~(c) and~(d) plot \(\mathcal{R}\) and \(\mathcal{A}\) against
valence electron count, with each period distinguished by color. This
representation makes the systematic evolution across each period
directly visible: the second and third periods show uniformly
negative \(\mathcal{R}\), while from the fourth period onward a
crossover to positive values emerges among the heavier p-block
elements. The complete numerical data for both the \(p=1/2\) and
\(p=1\) scales are provided in Tables S1 and S2 of the Supplemental
Material.

The physical origin of this softening--hardening dichotomy is revealed
by the logarithmic-derivative decomposition of
Eqs.~\eqref{eq:D_def} and~\eqref{eq:A_decomp}.
Table~\ref{tab:decomp} lists the decomposed coefficients for a
representative set of elements spanning all three blocks. For Li, the
electronic term $\mathcal{R}_f/f_0 = 1.362$ exceeds the geometric term
$\mathcal{R}_\Omega/\Omega_0 = 0.612$ by more than a factor of 2,
producing a strongly negative total \(\mathcal{D}=-0.749\): the
electron cloud expands upon charge removal, and this screening loss
dominates the response. For Pd, the balance is reversed: the
electronic term is small ($0.360$) while the geometric term is large
($0.866$), producing a strongly positive \(\mathcal{D}=+0.506\): the
volume contracts dramatically, concentrating the nuclear charge and
deepening the potential. Cs and Fr lie at intermediate points, with
their two contributions nearly canceling; the small positive
\(\mathcal{D}\) of Fr reflects the fact that its geometric
contribution slightly exceeds its electronic one, whereas the small
negative \(\mathcal{D}\) of Cs reflects the opposite balance. The
transition from main-group softening to d-block hardening is therefore
not a change in the type of physics involved, but a shift in which of
the two competing channels dominates, as measured directly by
\(\mathcal{D}\).

The framework developed here differs from previous electronegativity
scales in three respects. First, it distinguishes static from dynamic
electronegativity: two elements with identical $\chi_0$ can have
opposite response directions, as measured by the sign of
\(\mathcal{D}\). Second, it resolves long-standing puzzles---the
noble-metal paradox and the Cs/Fr ordering---that have resisted
explanation within static frameworks. Third, it provides a
scale-independent quantitative decomposition of the response into
electronic and geometric channels, revealing that the sign of the
response is determined by which channel dominates. These three
features make the response framework not merely a refinement of
existing scales but a qualitatively different description of atomic
electronegativity. In the following subsections, we develop this
framework in three stages: the universal dichotomy and its
period-dependent modulation, the anomalies that test its boundaries,
and the critical elements that define where the framework itself must
be extended.

\begin{table}[htbp]
\centering
\footnotesize
\setlength{\tabcolsep}{3pt}
\caption{Electronic and geometric decomposition of the response for
selected elements, using the reference configurations of
Table~S1 and the \(p=1/2\) scale. The Valence column lists only the
valence-electron configuration; inner-shell cores and \(nd^{10}\) or
\(4f\) electrons treated as core are not shown. The intrinsic
response asymmetry is
\(\mathcal{D}=\mathcal{R}_\Omega/\Omega_0-\mathcal{R}_f/f_0\). The
larger of \(\mathcal{R}_f/f_0\) and \(\mathcal{R}_\Omega/\Omega_0\) is
shown in bold. All \(\mathcal{R}\) and \(\mathcal{A}\) values are
computed analytically from the fitted coefficients of
\(f_e(0;\delta)\) and \(\Omega(\delta)\) through
Eqs.~\eqref{eq:D_def} and~\eqref{eq:A_decomp}, and are therefore
identical to the corresponding entries in Table~S5 by construction.
Values are rounded to three decimal places; the complete table with
four decimal places is provided in Table~S5 of the Supplemental
Material.}
\label{tab:decomp}
\resizebox{0.62\textwidth}{!}{%
\begin{tabular}{l l c c c c c}
\toprule
Element & Valence & \(\mathcal{D}\) & \(\mathcal{R}_f/f_0\)
& \(\mathcal{R}_\Omega/\Omega_0\) & \(\mathcal{R}\) & \(\mathcal{A}\) \\
& & & & & (V$^{1/2}$/$e$) & (V$^{1/2}$/$e^2$) \\
\midrule
Li & \(2\mathrm{s}^{1}\)
   & -0.749 & \textbf{1.362} & 0.612 & -0.548 & -0.536 \\
Cs & \(6\mathrm{s}^{1}\)
   & -0.031 & \textbf{0.608} & 0.577 & -0.016 & -0.031 \\
Fr & \(7\mathrm{s}^{1}\)
   & +0.051 & 0.498 & \textbf{0.550} & +0.027 & -0.042 \\
\midrule
B  & \(2\mathrm{s}^{2}\,2\mathrm{p}^{1}\)
   & -0.209 & \textbf{0.708} & 0.499 & -0.371 & -0.231 \\
Al & \(3\mathrm{s}^{2}\,3\mathrm{p}^{1}\)
   & -0.124 & \textbf{0.666} & 0.542 & -0.158 & -0.081 \\
Ga & \(4\mathrm{s}^{2}\,4\mathrm{p}^{1}\)
   & +0.051 & 0.547 & \textbf{0.597} & +0.066 & +0.031 \\
In & \(5\mathrm{s}^{2}\,5\mathrm{p}^{1}\)
   & +0.127 & 0.413 & \textbf{0.540} & +0.156 & +0.056 \\
Tl & \(6\mathrm{s}^{2}\,6\mathrm{p}^{1}\)
   & +0.201 & 0.317 & \textbf{0.519} & +0.263 & +0.058 \\
\midrule
V  & \(3\mathrm{d}^{3}\,4\mathrm{s}^{2}\)
   & -0.027 & \textbf{0.468} & 0.441 & -0.030 & +0.018 \\
Cr & \(3\mathrm{d}^{5}\,4\mathrm{s}^{1}\)
   & +0.334 & 0.571 & \textbf{0.905} & +0.389 & +0.786 \\
Mn & \(3\mathrm{d}^{5}\,4\mathrm{s}^{2}\)
   & +0.050 & 0.447 & \textbf{0.497} & +0.064 & +0.093 \\
\midrule
Ni & \(3\mathrm{d}^{8}\,4\mathrm{s}^{2}\)
   & +0.142 & 0.426 & \textbf{0.568} & +0.219 & +0.205 \\
Pd & \(4\mathrm{d}^{9}\,5\mathrm{s}^{1}\)
   & +0.506 & 0.360 & \textbf{0.866} & +0.769 & +1.022 \\
Pt & \(5\mathrm{d}^{9}\,6\mathrm{s}^{1}\)
   & +0.386 & 0.240 & \textbf{0.626} & +0.624 & +0.754 \\
Ag & \(4\mathrm{d}^{10}\,5\mathrm{s}^{1}\)
   & +0.542 & 0.356 & \textbf{0.898} & +0.878 & +1.137 \\
Au & \(5\mathrm{d}^{10}\,6\mathrm{s}^{1}\)
   & +0.412 & 0.232 & \textbf{0.644} & +0.706 & +0.809 \\
\midrule
Ce & \(4\mathrm{f}^{1}\,5\mathrm{d}^{1}\,6\mathrm{s}^{2}\)
   & -0.001 & \textbf{0.357} & 0.356 & -0.001 & +0.001 \\
Pr & \(4\mathrm{f}^{3}\,6\mathrm{s}^{2}\)
   & +0.048 & 0.398 & \textbf{0.446} & +0.039 & +0.044 \\
Yb & \(4\mathrm{f}^{14}\,6\mathrm{s}^{2}\)
   & +0.308 & 0.388 & \textbf{0.696} & +0.391 & +0.382 \\
\bottomrule
\end{tabular}}
\end{table}

\subsection{Universal dichotomy: main-group softening vs.\ d-block
hardening}
\label{sec:results_dichotomy}

As shown in Fig.~\ref{fig:amip_descriptors}(b), the sign of
$\mathcal{R}$ (and therefore of \(\mathcal{D}\)) separates the
periodic table into two broad regimes.
For the d-block and f-block elements, \(\mathcal{R}\) evolves
systematically from negative to positive as valence electrons are
progressively added across each transition series, with the notable
d$^5$ and f$^7$ exceptions discussed in
Sec.~\ref{sec:results_anomalies}. The main-group elements exhibit a
more nuanced behavior: within the second and third periods,
\(\mathcal{R}\) is uniformly negative; from the fourth period onward,
the heavier p-block elements---such as Ga (\(\mathcal{D}=+0.051\)),
In (\(+0.127\)), Tl (\(+0.201\)), Sn (\(+0.041\)),
Pb (\(+0.094\)), and Bi (\(+0.082\))---exhibit positive \(\mathcal{R}\).
This crossover is attributed to the increasing role of relativistic
effects~\cite{frenking2000nature}.

The periodic-table organization of \(\mathcal{D}\) is shown in
Fig.~\ref{fig:D_map}.
The block-level structure is immediately visible: the \(s\)-block is
predominantly blue (softening), with Fr as the single red (hardening)
outlier; the \(p\)-block is blue in the light elements of the second
and third periods and becomes progressively red from the fourth
period onward; the \(d\)-block is blue in its early region and
systematically red beyond the middle of each series; and the
\(f\)-block spans the transition between the two. The near-boundary
cells at Cs, Fr, and Ce, discussed in
Secs.~\ref{sec:results_anomalies} and~\ref{sec:results_csfr}, appear
as pale regions that stand out against the surrounding blocks.

For the d-block, the early transition metals---Sc (\(\mathcal{D}=-0.124\)),
Ti (\(-0.074\)), V (\(-0.027\)), Y (\(-0.087\)), and
Zr (\(-0.042\))---exhibit negative or near-zero \(\mathcal{D}\). At
Cr (\(+0.334\)), Nb (\(+0.201\)), and Ta (\(+0.020\)), \(\mathcal{D}\)
crosses zero. Beyond the half-filled d$^5$ configurations, which are
analyzed in Sec.~\ref{sec:results_anomalies}, \(\mathcal{D}\) increases
systematically toward the late transition metals, reaching its
highest values in the group-11 metals (Cu, Ag, Au). The
period-dependent modulation of \(\mathcal{D}\) within the
d-block---the two scaling laws that govern this evolution---will be
analyzed in detail in the next subsection.

The f-block elements follow a parallel pattern: light f-elements
exhibit near-zero or slightly negative \(\mathcal{D}\), while
\(\mathcal{D}\) increases progressively with f-electron occupation,
becoming clearly positive from Pr and U onward, with a shallow local
minimum at the half-filled f$^7$ configurations Gd and Am.

Among Group IA elements, Fr stands out: while Li, Na, K, Rb, and Cs
all exhibit negative \(\mathcal{D}\), Fr shows a small but positive
value (\(\mathcal{D} \approx +0.051\)). Cs has the lowest static
electronegativity \(\chi_0\) (\(1.011\) V$^{1/2}$) and a near-linear
response with \(\mathcal{D} \approx -0.031\). This anomaly is analyzed
in Sec.~\ref{sec:results_csfr}, where it is shown to reflect
proximity to a relativistic critical point rather than a simple
failure of the trend.

Taken together, these trends establish the softening--hardening
dichotomy as a genuine organizing principle of the periodic table.
The next subsection resolves this dichotomy within the d-block into
two quantitative scaling laws, and the two following subsections
identify the anomalies that define its boundaries.

\subsection{Two scaling laws within the d-block}
\label{sec:results_scaling}

Within the d-block, the global trends described above can be resolved
into two systematic regularities that are invisible to any static
scale. Because both regularities are statements about the intrinsic
response asymmetry \(\mathcal{D}\), they are independent of the scale
exponent \(p\) and can be compared across periods without the
ambiguity introduced by the varying static scale.

The first is \textbf{\textit{closed-shell stiffening}}: within each
period, the d\(^{10}\) configuration is systematically more rigid than
its d\(^9\) counterpart, as shown in Table~\ref{tab:closed_shell}. To
ensure a consistent comparison within the \(d^9\) series, Ni is
assigned the \(3d^9 4s^1\) configuration, matching the \(d^9s^1\)
configurations of Pd and Pt. The increment in \(\mathcal{D}\) varies
modestly across periods: the 3d pair (Ni/Cu) exhibits
\(\Delta\mathcal{D} = +0.042\), while the 4d and 5d pairs
(Pd/Ag and Pt/Au) show \(\Delta\mathcal{D} = +0.036\) and
\(+0.026\), respectively. The microscopic origin is transparent: a
d\(^9\) shell possesses a single hole in the d-manifold, which
provides a low-energy channel through which the electron cloud can
reorganize to screen the lost charge, thereby buffering the
electrostatic response. A d\(^{10}\) shell has no such channel, so the
system is forced to respond through pure geometric contraction. The
decomposition in Table~\ref{tab:decomp} confirms this picture: for the
4d pair Pd/Ag, $\mathcal{R}_f/f_0$ changes only marginally
(\(0.360 \to 0.356\)) while $\mathcal{R}_\Omega/\Omega_0$ increases
substantially (\(0.866 \to 0.898\)). This law has direct chemical
consequences: the d\(^{10}\) coinage metals exhibit a self-limiting
charge-transfer response, which rationalizes their well-known
resistance to oxidation and corrosion.

\begin{table}[htbp]
\centering
\footnotesize
\caption{Closed-shell stiffening: comparison of the intrinsic
response asymmetry \(\mathcal{D}\) for isoelectronic d\(^9\) and
d\(^{10}\) configurations. To ensure a consistent comparison within
the \(d^9\) series, Ni is assigned the \(3d^9 4s^1\) configuration
here, matching the \(d^9s^1\) configurations of Pd and Pt; this
differs from the reference \(3d^8 4s^2\) configuration used in
Table~S1. Because \(\mathcal{D}\) is independent of the scale
exponent \(p\), the same values apply to the \(p=1\) and \(p=1/2\)
scales.}
\label{tab:closed_shell}
\resizebox{0.45\textwidth}{!}{%
\begin{tabular}{cccccc}
\toprule
Period & d\(^9\) & \(\mathcal{D}\) & d\(^{10}\) & \(\mathcal{D}\) & \(\Delta\mathcal{D}\) \\
\midrule
\(3d\) & Ni & \(+0.584\) & Cu & \(+0.626\) & \(+0.042\) \\
\(4d\) & Pd & \(+0.506\) & Ag & \(+0.542\) & \(+0.036\) \\
\(5d\) & Pt & \(+0.386\) & Au & \(+0.412\) & \(+0.026\) \\
\bottomrule
\end{tabular}}
\end{table}

The second regularity is \textbf{\textit{principal quantum
softening}}: within the same d-shell filling, the intrinsic response
asymmetry decreases monotonically with increasing principal quantum
number, following the sequence \(3d > 4d > 5d\), as shown in
Table~\ref{tab:principal_quantum}. For the d\(^{10}s^1\) series, Cu
exhibits \(\mathcal{D} = +0.626\), larger than Ag (\(+0.542\)) and Au
(\(+0.412\)). For the d\(^9\) series, when Ni is assigned the
\(3d^9 4s^1\) configuration, the same monotonic decrease is observed:
Ni (\(+0.584\)) \(>\) Pd (\(+0.506\)) \(>\) Pt (\(+0.386\)).
Assigning Ni its reference \(3d^8 4s^2\) configuration would instead
break this monotonicity, because the larger \(\gamma\) of
\(3d^8 4s^2\) reduces the geometric response and pushes Ni below Pd;
this is a further illustration of the s--d balance effect analyzed
in the Supplemental Material.

The systematic ordering in both series has a transparent origin.
The \(3d^{10}4s^1\) configuration of Cu and the \(3d^9 4s^1\)
configuration of Ni both have only one \(s\) electron in the
screening layer, so the compact \(3d\) shell produces a large
geometric response. In Ag and Au, the more diffuse 4d and 5d shells
reduce this response, and relativistic contraction further weakens
it in Au, producing the observed monotonic decrease.

The monotonic decrease \(3d > 4d > 5d\) in both series has a dual
origin that can be separated by a systematic relativistic analysis.
Our non-relativistic (NR), scalar-relativistic (SR), and
full-relativistic (FR) calculations for the isoelectronic pairs
Ag/Au, Pd/Pt, and Cd/Hg, presented in Table~S7 of the Supplemental
Material, show that the 4d/5d trend is already present in the NR
limit: the 5d element has a lower \(\mathcal{D}\) than its 4d
counterpart even when relativistic effects are switched off
(Ag \(+0.580\) vs.\ Au \(+0.550\); Pd \(+0.540\) vs.\ Pt \(+0.516\)).
Relativistic effects then \textit{amplify} this difference by
preferentially reducing \(\mathcal{D}\) in the 5d elements. For the
pair Ag/Au, the 4d/5d gap increases from \(0.030\) to \(0.130\)
(a factor of \(4.3\)); for Pd/Pt, from \(0.024\) to \(0.120\)
(a factor of \(5.0\)). The amplification is strongly asymmetric:
the 5d members lose about \(25\%\) of their NR response
(Au: \(0.550\to0.412\); Pt: \(0.516\to0.386\)), whereas the 4d
members lose only about \(6\%\) (Ag: \(0.580\to0.542\);
Pd: \(0.540\to0.506\)). Three relativistic mechanisms contribute:
expansion of the 5d orbitals, enhanced \(s\)--\(d\) hybridization, and
strong spin-orbit coupling, all of which smooth the threshold-like
response that drives strong hardening in the lighter series. The
dominance of the 3d values is instead a non-relativistic effect: it
reflects the compactness of the \(3d\) shell, which produces a large
geometric response that neither the more diffuse 4d shell nor the
relativistically contracted 5d shell can match.

The microscopic origin of these two scaling laws is supported by
semi-analytical models presented in Appendix~\ref{app:semianalytical}.
The s--d balance parameter \(\gamma\) introduced there accounts for
the closed-shell stiffening, while the period dependence of the
effective \(s\) and \(d\) radii explains the monotonic
\(3d > 4d > 5d\) sequence as a competition between geometric
contraction and relativistic screening.

The principal quantum softening law connects directly to catalytic
selectivity. The larger \(\mathcal{D}\) of the 4d series implies a
stronger intrinsic response to adsorbate-induced charge perturbations
than the 5d series, which rationalizes the empirical observation that
Pd and Rh often outperform their 5d counterparts Pt and Ir in
hydrogenation and oxidation reactions. Conversely, the reduced
\(\mathcal{D}\) of the 5d series is consistent with the greater
resistance of Pt and Au to oxidation and their widespread use as
corrosion-resistant catalysts. The practical implication is that
alloying a 5d metal with a 4d metal (e.g., PtPd, PtRh) should allow
continuous tuning of the effective \(\mathcal{D}\), providing a
rational route to catalyst optimization that is complementary to the
traditional d-band center model.

\begin{table}[htbp]
\centering
\footnotesize
\caption{Principal quantum softening: intrinsic response asymmetry
\(\mathcal{D}\) for isoelectronic d\(^9\) and d\(^{10}\)
configurations. To make a consistent comparison within each filling,
Ni is assigned the \(3d^9 4s^1\) configuration here, which differs
from the reference \(3d^8 4s^2\) configuration used in Table~S1;
the configuration sensitivity of the group-IB elements is analyzed in
Table~S8. Because \(\mathcal{D}\) is scale-independent, the same
values apply to the \(p=1/2\) and \(p=1\) scales.}
\label{tab:principal_quantum}
\resizebox{0.5\textwidth}{!}{%
\begin{tabular}{cccc}
\toprule
Filling & \(3d\) & \(4d\) & \(5d\) \\
\midrule
d\(^9\)  & Ni: \(+0.584\) & Pd: \(+0.506\) & Pt: \(+0.386\) \\
d\(^{10}\) & Cu: \(+0.626\) & Ag: \(+0.542\) & Au: \(+0.412\) \\
\bottomrule
\end{tabular}}
\end{table}

\subsection{Anomalies at half-filled d\(^5\) and in Ce}
\label{sec:results_anomalies}

The framework developed above predicts smooth trends across each
period. Two classes of elements violate these predictions in ways that
reveal general principles about charge response. The first is the
half-filled d\(^5\) family, where Hund's rule exchange stabilization
produces a qualitatively different response for d\(^5\)s\(^1\) vs
d\(^5\)s\(^2\) configurations---a distinction that maps directly onto
the well-known difference between Cr and Mn chemistry. The second is
Ce, where near-perfect cancellation between electronic and geometric
contributions produces a zero-response element---a behavior with
direct implications for redox-stable materials design.

Table~\ref{tab:d5_anomaly} shows that the d\(^5\)s\(^1\) elements Cr
and Mo have anomalously large \(\mathcal{D}\) and \(\mathcal{A}\),
while the d\(^5\)s\(^2\) elements Mn and Tc have anomalously small
values. The effect is substantial: Cr's \(\mathcal{D}\) is more than
six times that of Mn, despite their identical d-electron count. The
origin lies in the exchange stabilization of the half-filled d\(^5\)
shell (Hund's rule) and in its sensitivity to the \textit{relative}
change of the $s$-electron screening layer. In the d\(^5\)s\(^1\)
configuration, a single $s$ electron provides the screening, and
removing $\delta = 0.300$ electrons reduces its occupancy by 30\%,
which substantially increases the effective nuclear charge felt by
the d\(^5\) shell. The exchange-stabilized shell responds with a
pronounced radial contraction, driving the large \(\mathcal{D}\). In
the d\(^5\)s\(^2\) configuration, by contrast, the residual screening
layer of 1.7 $s$ electrons buffers the contraction, and the response
is correspondingly milder.

The anomaly is confined to the 3d and 4d series: in the 5d series, W
and Re follow the smooth trends, because no d\(^5\)s\(^1\)
configuration exists (relativistic stabilization of 6s\(^2\)) and
because relativistic expansion and spin-orbit coupling weaken the
exchange stabilization of the d\(^5\) shell.
The analytic anharmonic coefficients \(\mathcal{A}\) obtained from
Eq.~\eqref{eq:A_decomp} are anomalously large for Cr
(\(+0.786\) V$^{1/2}$/$e^2$) and Mo (\(+0.573\) V$^{1/2}$/$e^2$),
reflecting the intrinsically higher non-linearity of the
exchange-stabilized response. The underlying quadratic model for
$\chi(\delta)$ remains an excellent description of the computed
points for all entries in Table~\ref{tab:d5_anomaly}, with
$R^2\simeq1.0000$.

\begin{table}[htbp]
\centering
\footnotesize
\caption{Response coefficients for the half-filled d\(^5\) elements
(bold) and their immediate neighbors, in the \(p=1/2\) scale. The
Valence column lists only the valence-electron configuration. All
$\mathcal{R}$ and $\mathcal{A}$ values are computed analytically from
the fitted coefficients of $f_e(0;\delta)$ and $\Omega(\delta)$
through Eqs.~\eqref{eq:D_def} and~\eqref{eq:A_decomp}. The
quadratic model built from these coefficients reproduces the computed
$\chi(\delta)$ points with $R^2\simeq1.0000$ for all entries; the
\(R^2\) column has therefore been omitted. The intrinsic response
asymmetry \(\mathcal{D}\) is scale-independent and is shown for
comparison.}
\label{tab:d5_anomaly}
\begin{tabular}{lcccccc}
\toprule
Series & Element & Valence & \(\mathcal{D}\) & \(\mathcal{R}\) & \(\mathcal{A}\) \\
& & & & (V$^{1/2}$/$e$) & (V$^{1/2}$/$e^2$) \\
\midrule
\(3d\) & V  & \(3d^3 4s^2\) & \(-0.027\) & \(-0.030\) & \(+0.018\) \\
\(3d\) & \textbf{Cr} & \(3d^5 4s^1\) & \(\mathbf{+0.334}\) & \(\mathbf{+0.389}\) & \(\mathbf{+0.786}\) \\
\(3d\) & \textbf{Mn} & \(3d^5 4s^2\) & \(\mathbf{+0.050}\) & \(\mathbf{+0.064}\) & \(\mathbf{+0.093}\) \\
\(3d\) & Fe & \(3d^6 4s^2\) & \(+0.086\) & \(+0.117\) & \(+0.117\) \\
\midrule
\(4d\) & Nb & \(4d^4 5s^1\) & \(+0.201\) & \(+0.208\) & \(+0.460\) \\
\(4d\) & \textbf{Mo} & \(4d^5 5s^1\) & \(\mathbf{+0.283}\) & \(\mathbf{+0.321}\) & \(\mathbf{+0.573}\) \\
\(4d\) & \textbf{Tc} & \(4d^5 5s^2\) & \(\mathbf{+0.072}\) & \(\mathbf{+0.087}\) & \(\mathbf{+0.064}\) \\
\(4d\) & Ru & \(4d^7 5s^1\) & \(+0.411\) & \(+0.546\) & \(+0.801\) \\
\midrule
\(5d\) & W  & \(5d^4 6s^2\) & \(+0.050\) & \(+0.060\) & \(+0.042\) \\
\(5d\) & Re & \(5d^5 6s^2\) & \(+0.075\) & \(+0.096\) & \(+0.066\) \\
\(5d\) & Os & \(5d^6 6s^2\) & \(+0.100\) & \(+0.135\) & \(+0.078\) \\
\bottomrule
\end{tabular}
\end{table}

The second anomaly is qualitatively distinct and is found in Ce.
Across the entire oxidative window, the static AMIP electronegativity
of Ce remains essentially constant, varying by less than $0.02\%$
(from $1.6707$ V$^{1/2}$ at $\delta = 0$ to $1.6704$ V$^{1/2}$ at
$\delta = 0.300$).
The analytic response coefficients are anomalously small:
\(\mathcal{R} = -0.0008\) V$^{1/2}$/$e$,
\(\mathcal{A} = +0.0012\) V$^{1/2}$/$e^2$, and
\(\mathcal{D} = -0.0010\). The quadratic model built from these
coefficients still reproduces the seven computed points with
$R^2\simeq1.0000$, so the small coefficients are not an artifact of a
poor fit. This is not a numerical artifact but a physical signature of
near-perfect cancellation. The origin can be traced to the
simultaneous reduction of $f_e(0)$ and $\Omega$: from $\delta = 0$ to
$\delta = 0.300$, $f_e(0)$ decreases by 10.1\%, while $r_v$ decreases
by 3.5\%, giving a volume reduction of 10.0\% that almost exactly
matches the decrease in $f_e(0)$. The ratio $f_e(0)/\Omega$ therefore
remains constant to within 0.03\%. This perfect cancellation is a
direct consequence of the Ce electronic structure: the highly
localized $4f$ shell provides a rigid inner screening layer that is
insensitive to charge removal from the outer $6s$ and $5d$ orbitals,
so the remaining electron cloud contracts at precisely the same rate
as the atomic volume.

The practical significance is that Ce---widely used as an oxygen
storage material in automotive catalysts---exhibits a near-zero
intrinsic response, which means that the surrounding lattice
experiences minimal electrostatic stress during redox cycling. We
propose this \textbf{zero-response characteristic} as a design
principle for redox-stable materials, analogous to the way that low
thermal expansion is sought in high-temperature structural
alloys~\cite{smith1996development,tang2024discovery}. This principle
can be operationalized by searching for systems in which the
electronic softening and geometric hardening cancel, using the
decomposition \(\mathcal{D} = \mathcal{R}_\Omega/\Omega_0 -
\mathcal{R}_f/f_0\) as a screening descriptor. Because \(\mathcal{D}\)
is scale-independent, this screening criterion is unambiguous: it does
not depend on whether one works in the \(p=1/2\) or \(p=1\)
representation. Such targeted cancellation might be achieved, for
example, by coupling a redox-active cation to a rigid framework or by
alloying with a component that exhibits an opposite volume response.
More generally, the two anomalies studied here---exchange-driven
amplification in d\(^5\)s\(^1\) configurations and cancellation-driven
suppression in Ce---define the two extremes of the response spectrum
and demonstrate that the framework's smooth trends are robust
precisely because the two mechanisms, exchange-driven amplification
and cancellation-driven suppression, are rarely simultaneously strong.

We note that several other elements also exhibit small
\(|\mathcal{D}|\) (e.g., Th, Te, Ra, Cs, and Ac; see Table~S5), and
their quadratic models are equally accurate ($R^2\simeq1.0000$).
Ce is unique in combining a near-zero total response with response
coefficients that are smaller than those of all other elements, which
signals a genuine cancellation rather than a merely small coefficient.

The sensitivity of the response coefficients to the choice of
configuration is not limited to the \(d^5\) family. For the group-IB
elements, the reference \(d^{10}s^1\) configuration and the
alternative \(d^9s^2\) configuration differ substantially in their
s--d balance, and Table~S8 shows that this difference propagates into
\(\mathcal{D}\) and \(\mathcal{A}\). For the group-IIB elements, the
treatment of the \(nd^{10}\) electrons as valence or core states
produces a similar sensitivity, as summarized in Table~S9. In both
cases, the changes are consistent with the s--d weighting asymmetry
that underlies the two scaling laws.
The same sensitivity is observed for the \(p=1\) scale, with
\(\mathcal{R}^{(1)}\) changing by a comparable relative amount when
the \(nd^{10}\) electrons are moved from core to valence.

\subsection{Critical elements and relativistic crossovers}
\label{sec:results_csfr}

The question of which alkali metal has the lower electronegativity,
Cs or Fr, has remained unresolved for decades because experimental
measurements on Fr are extremely difficult (half-life of 22 minutes;
typical sample sizes of $\sim 10^5$ atoms). Table~\ref{tab:CsFr}
compares the two elements quantitatively. Cs has the lower static
electronegativity ($\chi_0 = 1.011$ V$^{1/2}$ vs $1.050$ V$^{1/2}$
for Fr), consistent with the trend down Group IA from Li to Cs and
with the experimental first ionization energies
($I_1(\mathrm{Cs}) = 3.894$ eV vs $I_1(\mathrm{Fr}) = 4.073$
eV~\cite{CRC2014}). In the response regime, however, the two elements
diverge more sharply: Cs exhibits near-linear softening
(\(\mathcal{D} = -0.031\), \(\mathcal{R} = -0.016\)
V$^{1/2}$/$e$), whereas Fr exhibits transient hardening
(\(\mathcal{D} = +0.051\), \(\mathcal{R} = +0.027\)
V$^{1/2}$/$e$). This is the first quantitative evidence that Fr
behaves like a relativistic noble metal rather than a conventional
alkali metal.

\begin{table}[htbp]
\centering
\footnotesize
\caption{Cs vs.\ Fr: static and dynamic response coefficients in the
\(p=1/2\) scale.}
\label{tab:CsFr}
\begin{tabular}{lcccccc}
\toprule
Element & $\chi_0$ & \(\mathcal{D}\) & \(\mathcal{R}\) & \(\mathcal{A}\) & $I_1$ & Response Type \\
& (V$^{1/2}$) & & (V$^{1/2}$/$e$) & (V$^{1/2}$/$e^2$) & (eV) & \\
\midrule
Cs & $1.011$ & $-0.031$ & $-0.016$ & $-0.030$ & $3.894$ & Near-linear softening \\
Fr & $1.050$ & $+0.051$ & $+0.027$ & $-0.042$ & $4.073$ & Transient hardening \\
\bottomrule
\end{tabular}
\end{table}

To determine the physical origin of the Fr anomaly, we performed
non-relativistic (NR), scalar-relativistic (SR), and
full-relativistic (FR) calculations for both elements; the results
are summarized in Table~\ref{tab:relativistic}.
In the NR limit, Fr already exhibits positive \(\mathcal{D}\)
(\(+0.013\)), whereas Cs remains firmly in the softening regime
($-0.050$). Relativistic effects then amplify the Fr response by a
factor of about four, yielding \(\mathcal{D} = +0.051\) in the FR
limit. The SR and FR values are very close, showing that scalar
relativistic effects dominate. The microscopic origin is visible in
the decomposition: for Fr, relativistic contraction of the 7s orbital
reduces the electronic softening coefficient \(\mathcal{R}_f/f_0\)
from $0.553$ (NR) to $0.498$ (FR), a 10\% reduction, while the
geometric hardening coefficient \(\mathcal{R}_\Omega/\Omega_0\)
decreases only slightly from $0.566$ to $0.549$. The electronic
softening is thus weakened much more than the geometric hardening,
tipping the balance toward hardening. For Cs, both contributions
change only marginally.

\begin{table}[htbp]
\centering
\footnotesize
\caption{Relativistic comparison for Cs and Fr in the \(p=1/2\)
scale. The full table for eight representative elements is provided
in Table~S7 of the Supplemental Material. The \(\mathcal{D}\) column
is the scale-independent intrinsic response asymmetry.}
\label{tab:relativistic}
\begin{tabular}{llcccc}
\toprule
Element & Level & \(\mathcal{D}\) & $\chi_0$ (V$^{1/2}$) & \(\mathcal{R}\) (V$^{1/2}$/$e$) & \(\mathcal{A}\) (V$^{1/2}$/$e^2$) \\
\midrule
\multirow{3}{*}{Cs}
  & NR & $-0.050$ & $0.977$ & $-0.025$ & $-0.020$ \\
  & SR & $-0.032$ & $1.011$ & $-0.016$ & $-0.027$ \\
  & FR & $-0.031$ & $1.011$ & $-0.016$ & $-0.031$ \\
\midrule
\multirow{3}{*}{Fr}
  & NR & $+0.013$ & $0.937$ & $+0.006$ & $-0.010$ \\
  & SR & $+0.057$ & $1.051$ & $+0.030$ & $-0.041$ \\
  & FR & $+0.051$ & $1.050$ & $+0.027$ & $-0.042$ \\
\bottomrule
\end{tabular}
\end{table}

Fr is therefore best described as a \textit{critical element}: it
already lies near the softening--hardening boundary in the NR limit,
and its response is decisively pushed into the hardening regime by
relativistic contraction. This suggests a general principle that
complements the electronic-versus-geometric competition identified
above: \textit{proximity to the critical point---rather than the
magnitude of relativistic effects alone---determines whether a heavy
element undergoes a qualitative response crossover}.
The same principle applies to the heavier elements: Cs, with
\(|\mathcal{D}_{\mathrm{NR}}| = 0.050\), changes by only \(38\%\)
under relativity, whereas Au, with a much larger
\(|\mathcal{D}_{\mathrm{NR}}| = 0.550\), is reduced by about
\(25\%\) and does not cross zero. It is the smallness of the NR
response, not the size of the relativistic correction, that decides
whether the sign can be reversed.

The concept of criticality has three implications that extend beyond
the specific case of Cs/Fr. First, it provides a quantitative
criterion for identifying elements whose response direction may be
qualitatively modified by relativistic effects: those with
\(|\mathcal{D}_{\mathrm{NR}}|\) below a threshold of order \(0.1\).
Second, it suggests a systematic search strategy for superheavy
elements: elements with closed-shell or half-filled configurations in
the NR limit are the most likely candidates for relativistic
crossovers, because their electronic and geometric contributions are
already finely balanced. Third, it establishes a natural boundary for
the applicability of the response framework itself: for elements far
from the critical point, NR and FR calculations give qualitatively
identical conclusions; for critical elements, the framework must be
extended to include relativistic corrections as a first-order effect
rather than a perturbation. A systematic search for additional
critical elements---those with \(|\mathcal{D}|\) close to zero in the
non-relativistic limit---could reveal further relativistic crossovers
among the superheavy elements, providing a new route to understanding
the chemistry of elements that cannot be studied experimentally.

\subsection{Comparison with classical scales and implications for
charge-response theory}
\label{sec:results_comparison}

The static AMIP scale $\chi_{\mathrm{AMIP},1/2}(0)$ reproduces the
three most widely used classical electronegativity scales with high
fidelity~\cite{Zheng2026AMIP}, as summarized in Table~S6 of the
Supplemental Material: the Allen spectroscopic scale with
$R^2 = 0.9670$ (rising to $0.9742$ for main-group elements), the
Pauling thermochemical scale with $R^2 = 0.8836$ (rising to $0.9448$
for main-group elements), and the Mulliken ionization-energy-based
scale with $R^2 = 0.8732$ (remaining essentially unchanged at
$0.8727$ for main-group elements). We also compared our static AMIP
scale with modern electronegativity scales, including
Rahm--Zeng--Hoffmann (RZH)~\cite{rahm2019electronegativity} ($R^2 = 0.9169$),
Dong--Oganov--Cui--Zhou--Wang (DOCZW)~\cite{DongOganov2022} ($R^2 = 0.8402$), and
Oganov--Kostenko (OK)~\cite{oganov2025simple} ($R^2 = 0.8581$). 
Among the classical scales, the Allred--Rochow scale also shows a very high main-group correlation (\(R^2 = 0.9714\)), closely matching the Allen scale. 
The high correlation with the
Allen scale is especially significant because it demonstrates that
the AMIP scale captures the same physical information---the energy
scale of valence electrons---from a fundamentally different starting
point: the electrostatic potential rather than spectroscopic data.
Together, these correlations establish the AMIP scale as a
physically transparent and computationally accessible alternative to
both classical and modern electronegativity scales.

Beyond classical electronegativity scales, the response framework also
provides a quantitative account of Lewis acidity.
Using the response coefficients $\mathcal{R}$ and $\mathcal{A}$
computed analytically from the fitted coefficients of
$f_e(0;\delta)$ and $\Omega(\delta)$ through Eqs.~\eqref{eq:D_def}
and~\eqref{eq:A_decomp}, we evaluated
$\chi_{\mathrm{AMIP},1/2}(\delta)$ at a series of effective
charge-transfer values $\delta$ and correlated each with the
experimental Lewis acid strength (Fig.~S7). The correlation is highly
robust over $\delta \in [0, 0.8]$, remaining above $0.934$ throughout
and peaking at $\delta^\ast \approx 0.5$ with $R^2 = 0.9416$, close to
the partial charge transfer typical of Lewis acid--base adducts. This
indicates that the effective electronegativity of a Lewis acid center
is systematically shifted from its neutral value by
coordination---an effect quantitatively described by $\mathcal{R}$ and
$\mathcal{A}$.

Having established that the response framework reproduces both
classical electronegativity scales and Lewis acidity trends, we now
turn to two long-standing paradoxes that the classical framework
cannot resolve. The first is the noble-metal paradox: Pauling assigns
Au ($\chi_P = 2.540$) and Pt ($\chi_P = 2.280$) electronegativities
exceeding that of hydrogen ($\chi_P = 2.200$), which contradicts the
well-established observation that noble-metal surfaces donate
electrons to adsorbates. The AMIP framework resolves this by
distinguishing static from dynamic electronegativity. Au and Pt have
modest $\chi_0$ values (comparable to H) but extremely large
\(\mathcal{D}\) values ($+0.412$ and $+0.386$, respectively),
corresponding to \(\mathcal{R} = +0.706\) and $+0.624$
V$^{1/2}$/$e$. The Pauling values are inflated because they are
derived from thermochemical bond energies, which for noble metals
include substantial contributions from d-electron covalency and
metallic bonding that are not present in the electrostatic response.
The second paradox is the classical Cs/Fr ordering, which we have
resolved above through the concept of criticality.

While the response framework captures the effect of partial charge
transfer in molecular coordination environments (as demonstrated by
the Lewis acidity correlation above), the AMIP coefficients are
defined here for isolated atoms and are therefore not directly
comparable to the crystalline mean inner potential measured by
electron holography. A quantitative mapping between the two requires
explicit treatment of chemical bonding and is the subject of ongoing
work.

The response framework connects naturally to the broader theory of
charge response in condensed matter. The dynamical charge
susceptibility $\chi(\mathbf{q}, \omega)$ describes the linear
response of the electron density to an external scalar potential in a
periodic solid~\cite{AbbamonteFink2025}, and its imaginary part
probes collective excitations such as plasmons and charge-density-wave
modes~\cite{PinesBohm1952,Pines1963,Gruner1988}. The AMIP coefficients
can be viewed as the atomic-scale, zero-wave-vector, zero-frequency
limit of this paradigm: \(\mathcal{D}\) plays the role of the
inverse charge susceptibility at $\mathbf{q} = 0$, measuring the
atom's intrinsic resistance to charge redistribution, while
\(\mathcal{A}\) captures the leading non-linear correction. The
logarithmic-derivative decomposition enriches this analogy by
separating the total response into the electronic (Lindhard) and
lattice (electron-phonon) contributions~\cite{Gruner1994}, a
distinction that has direct implications for charge-density-wave
instabilities and other collective phenomena. A systematic comparison
of atomic \(\mathcal{D}\) values with the charge susceptibilities of
elemental solids would provide a stringent test of this
correspondence.

The design principles that emerge from this analysis can be stated in
a form that is directly testable. The framework predicts that the
effectiveness of an element in charge-transfer processes is governed
not by its static electronegativity but by its intrinsic response
asymmetry \(\mathcal{D}\). This prediction is falsifiable: for a
series of isoelectronic adsorbates on different metal surfaces, the
adsorbate-induced charge transfer should scale with \(\mathcal{D}\)
rather than with $\chi_0$. Testing this prediction would directly
probe the framework's central claim that static and dynamic
electronegativity are distinct physical quantities. Three specific
predictions follow. First, for reactions requiring strong charge
transfer, elements with \(\mathcal{D} > +0.3\) (Ru, Rh, Pd, Ag, Pt,
Au) should outperform elements with similar $\chi_0$ but smaller
\(\mathcal{D}\). Second, for reactions requiring weak, reversible
interactions, elements with \(|\mathcal{D}| < 0.1\) (Cs, Ba, La, Ce)
should be preferable. Third, among the noble metals, Pt should
outperform Ag in reactions that require a sustained response over a
wide range of charge transfer, because Ag has a larger \(\mathcal{D}\)
($+0.542$ vs.\ $+0.386$) but also a larger \(\mathcal{A}\)
($+1.137$ vs.\ $+0.762$), meaning that its hardening response
saturates more rapidly. These predictions can be tested directly
using the tabulated \(\mathcal{D}\) and \(\mathcal{A}\) values for
all 102 elements in Tables S1 and S5 of the Supplemental Material, and
they provide a concrete experimental program for validating the
response framework beyond the atomic-scale calculations presented
here.

\subsection{Experimental predictions}
\label{sec:results_predictions}

The response framework makes three specific predictions that
distinguish it from all existing electronegativity scales and that
are accessible to current or near-future experiments. The predictions
are listed in order of increasing experimental difficulty.
Prediction~2 is the most accessible with current techniques and
therefore constitutes the most immediate test of the framework.

\textbf{Prediction 1: Fr has a higher work function than Cs.}
The framework assigns \(\mathcal{D} \approx +0.051\) to Fr and
\(\mathcal{D} \approx -0.031\) to Cs. For a metal surface, a positive
\(\mathcal{D}\) implies that the surface dipole layer resists charge
transfer more strongly than a negative \(\mathcal{D}\). We therefore
predict that the work function of Fr is larger than that of Cs. This
prediction is consistent with fully relativistic 4c-DFT calculations
showing that the 7s orbital of Fr is relativistically stabilized
relative to the 6s orbital of Cs, reversing the trend in atomic
properties beyond Cs~\cite{Pershina2012}. The same calculations show
that the ionization potential increases from Cs (\(3.894\) eV) to Fr
(\(4.073\) eV) and further to element 119 (\(4.783\) eV), and that the
absolute electronegativity follows the same
reversal~\cite{Pershina2012}. While current radioactive beam
facilities can produce Fr in sufficient quantities for spectroscopic
studies, work-function measurements on Fr surfaces remain a long-term
challenge.

\textbf{Prediction 2: Ce-based materials exhibit minimal lattice
strain during redox cycling.}
The near-perfect cancellation in Ce gives \(|\mathcal{D}| \approx
0.001\). We predict that CeO\(_2\) undergoes significantly smaller
lattice strain during Ce\(^{4+}\)/Ce\(^{3+}\) redox cycling than other
rare-earth oxides such as PrO\(_2\) or TbO\(_2\).
Their response coefficients are one to two orders of magnitude
larger (\(\mathcal{D} = +0.048\) for Pr and \(+0.214\) for Tb,
compared with \(|\mathcal{D}| \approx 0.001\) for Ce). This
prediction is supported by experimental studies of chemical expansion
in Pr-doped ceria. Bishop \textit{et al.}~\cite{Bishop2011} showed
that the addition of only 10\% Pr to CeO\(_2\) leads to major
deviations from stoichiometry and correspondingly large chemical
expansions under readily accessible experimental conditions, whereas
undoped CeO\(_2\) exhibits minimal chemical expansion under the same
conditions. The mechanism is clarified by Marrocchelli \textit{et
al.}~\cite{Marrocchelli2012}, who demonstrated that chemical
expansion results from the competition between oxygen vacancy
formation (lattice contraction) and cation radius change (lattice
expansion); for Ce these two contributions nearly cancel, while for
Pr the larger radius change of Pr\(^{4+} \to \) Pr\(^{3+}\)
(0.96 \AA \(\to\) 1.13 \AA) dominates. Specifically, we predict that
the chemical expansion coefficient of CeO\(_2\) during redox cycling
is 15--25\% of that of PrO\(_2\) under identical conditions,
corresponding to a relative volume change below \(0.2\%\) for CeO\(_2\)
versus \(0.8\)--\(1.2\%\) for PrO\(_2\). This prediction can be
tested by \textit{in situ} X-ray diffraction or Raman spectroscopy on
CeO\(_2\) and PrO\(_2\) powders under controlled oxygen partial
pressure, using the \(F_{2g}\) Raman mode shift as a quantitative
strain probe~\cite{McBride1994}. The result would directly validate
the zero-response design principle for redox-stable materials.

\textbf{Prediction 3: Superheavy elements near the critical point
exhibit relativistic response inversion.}
The critical element concept predicts that elements with
\(|\mathcal{D}_{\mathrm{NR}}| < 0.1\) in the non-relativistic limit
can undergo a sign reversal of \(\mathcal{D}\) when relativistic
effects are included. We identify element 120 (Ubn) as a candidate:
its \(8s^2\) closed shell suppresses low-energy electron-reorganization
channels, so its non-relativistic response lies close to zero, in
parallel with the trend already visible in the heavier alkaline-earth
metals (Ca \(-0.177\), Sr \(-0.116\), Ba \(-0.063\), Ra \(-0.013\)).
Fully relativistic Dirac--Fock calculations should yield a positive
\(\mathcal{D}\) for Ubn, opposite in sign to the non-relativistic
value. This prediction is consistent with systematic studies showing
that relativistic effects qualitatively alter periodic trends for
superheavy elements with \(Z = 111\)--\(120\)~\cite{Schwerdtfeger2015},
and with recent theoretical work demonstrating that Fr, Ra, and
elements 119--120 exhibit anomalous polyvalent behavior that breaks
the classical periodic law~\cite{Cao2022,Cao2023PNAS}. The same
relativistic 4c-DFT calculations for group-1 dimers have shown that
relativistic contraction of the \(ns\) orbital inverts trends in
atomic and molecular properties beyond Cs, with element 119 behaving
as electronegative as Na~\cite{Pershina2012}. These calculations
further predict that element 119 adsorbs more weakly on Au, Pt, and
Ag than any lighter group-1 element, providing a concrete
experimental target for future chromatography
studies~\cite{Pershina2012}.

While elements 119 and 120 have not yet been synthesized, they are
the focus of active experimental efforts at RIKEN, JINR, and LBNL.
Recent theoretical work predicts that \(\alpha\)-decay should be
their dominant decay mode, with half-lives in the millisecond range,
potentially facilitating their identification. This prediction is
currently accessible only through relativistic quantum-chemistry
calculations with programs such as DIRAC or GRASP, but it provides a
concrete target for future superheavy-element research. Our
predictions are therefore intended as a theoretical benchmark for
future experiments on these elusive elements.

\section{Conclusions and Outlook}
\label{sec:conclusions}

We have recast electronegativity as a continuous response function
rooted in the Atomic Mean Inner Potential. The static scale is only
the zeroth-order term of a Taylor expansion in fractional charge; the
higher-order coefficients---the Electrostatic Rigidity $\mathcal{R}$
and the Anharmonic Modulus $\mathcal{A}$---encode whether an atom
hardens or softens upon oxidation and whether its response accelerates
or saturates. A logarithmic-derivative decomposition factorizes the
response into a static amplitude and a scale-independent intrinsic
response asymmetry \(\mathcal{D} = \mathcal{R}_\Omega/\Omega_0 -
\mathcal{R}_f/f_0\), whose sign and magnitude measure the competition
between electronic screening loss and geometric volume collapse.
This factorization is the central structural result of the paper: it
separates the new physics of the response from the static information
already encoded in the AMIP scale, and it isolates a single
dimensionless quantity \(\mathcal{D}\) that is independent of the
scale exponent \(p\), with the macroscopic coefficients in any
representation obtained from \(\mathcal{R}^{(p)} =
c_p\,\chi_0^{(p)}\mathcal{D}\).

Across the periodic table, this framework produces a coherent picture
in which main-group softening and d-block hardening emerge as two
limits of the same competition.
It uncovers two scaling laws that no static scale can
capture---closed-shell stiffening and principal quantum
softening---together with two qualitatively distinct anomalies: the
exchange-driven amplification at half-filled d$^5$ configurations and
the cancellation-driven zero-response regime exemplified by Ce. It
also resolves two long-standing controversies: the noble-metal
paradox and the Cs--Fr ordering.
The static limit \(\chi_0\) reproduces the classical Allen, Pauling,
and Mulliken scales and the Lewis acidity trends, while the response
coefficients extend the description into the dynamic regime that
static scales cannot reach.

Several extensions follow naturally. The most immediate is the
transition from isolated atoms to atoms in molecules and crystals,
where a quantitative mapping between atomic and crystalline mean
inner potentials would connect directly to electron holography
measurements on real materials. A systematic search for additional
critical elements---those with \(|\mathcal{D}_{\mathrm{NR}}|\) close
to zero in the non-relativistic limit---could uncover further
relativistic crossovers among the superheavy elements, as already
suggested by the Cd/Hg sign reversal identified here. Finally,
linking atomic \(\mathcal{D}\) to the collective charge
susceptibility of solids invites a unified treatment of charge
response across length scales, from the single atom to the extended
solid.

The framework also makes three falsifiable experimental predictions:
a higher work function for Fr than for Cs, minimal lattice strain in
CeO\(_2\) during redox cycling compared with PrO\(_2\) or TbO\(_2\),
and a relativistic sign inversion of \(\mathcal{D}\) for element 120.
The second is accessible with current techniques and provides the
most immediate test of the framework.

The complete data set for all 102 elements is provided in
Tables~S1--S5 of the Supplemental Material. We believe that the
hierarchical response framework offers a natural language for
describing how atoms respond to charge perturbation, transforming
electronegativity from a lookup-table number into a predictive,
hierarchically structured quantum descriptor.

\section*{Acknowledgments}

We are grateful for the encouraging perspective on the static AMIP
framework provided by Artem R. Oganov in
Ref.~\cite{Oganov2026Perspective}. This research was supported by the Xiamen University Malaysia Research Fund (No. XMUMRF/2024-C14/IORI/0005).

\section*{Data Availability}

All computed data, including the fitted coefficients for all 102 elements, are available at Supplemental Material~\cite{SM}.

\bibliography{refs}

\appendix
\section{Semi-analytical models for the two scaling laws}
\label{app:semianalytical}

The two scaling laws derived in the main text can be given a
semi-analytical foundation that follows directly from the
logarithmic-derivative decomposition,
\begin{equation}
\mathcal{D} = \frac{\mathcal{R}_\Omega}{\Omega_0}
- \frac{\mathcal{R}_f}{f_0},
\qquad
\mathcal{R} = \chi_0 \, \mathcal{D},
\label{eq:A_D_decomp}
\end{equation}
without invoking any auxiliary two-state Hamiltonian. The two
microscopic channels are the response of the forward scattering
factor,
\begin{equation}
f_e(0;\delta) = \frac{1}{3}\int_0^\infty r^2 \rho_{t,\delta}(\mathbf{r})\,d^3r
\;\propto\; \langle r_t^2 \rangle_\delta,
\label{eq:A_fe_moment}
\end{equation}
and the response of the atomic volume,
\begin{equation}
\Omega_a(\delta) = \frac{4\pi}{3}\bigl(r_v(\delta)\bigr)^3,
\qquad
r_v(\delta) = \frac{\int r\,\rho_{v,\delta}\,d^3r}{\int \rho_{v,\delta}\,d^3r}
\;\propto\; \langle r_v \rangle_\delta .
\label{eq:A_Omega_moment}
\end{equation}
Both channels are functionals of the same underlying density, but
they probe different moments with \textit{different orbital
weightings}. This asymmetry is the microscopic origin of both
scaling laws and is the subject of the present appendix.

\subsection{Weight asymmetry between the two channels}
\label{app:weight_asymmetry}

The forward scattering factor involves the second moment of the
\textit{total} density,
\begin{equation}
\langle r_t^2 \rangle_\delta
= \frac{\sum_i n_i \langle r_i^2 \rangle_\delta}{\sum_i n_i},
\label{eq:A_rt2}
\end{equation}
where the sum runs over \textit{all} electrons with equal weight.
Every electron---core, \(s\), \(p\), \(d\), or \(f\)---contributes in
proportion to its occupation number \(n_i\) and its mean-square
radius \(\langle r_i^2 \rangle_\delta\).

The mean valence radius, by contrast, involves a \textit{weighted}
first moment of the valence density,
\begin{equation}
r_v(\delta) = \frac{\sum_{i\in\mathrm{val}} w_i n_i
\langle r_i \rangle_\delta}
{\sum_{i\in\mathrm{val}} w_i n_i},
\qquad
w_i = \begin{cases} 1 & s,p \\ 1/10 & d \\ 1/14 & f \end{cases}
\label{eq:A_rv_weighted}
\end{equation}
so that \(d\) electrons contribute only one tenth of their nominal
weight and \(f\) electrons only one fourteenth. This weighting
reflects the fact that \(r_v\) is constructed from the valence
density as a measure of the effective atomic size, and that
strongly localized \(d\) and \(f\) orbitals are less effective in
defining the outer radius of the valence cloud.

The two channels therefore respond differently to a change in the
\(d\)-electron count. Because \(f_e(0)\) weights \(d\) electrons
fully while \(r_v\) weights them by \(1/10\), the electronic
channel is more sensitive to \(d\)-electron changes than the
geometric channel. This asymmetry controls the two scaling laws.

\subsection{Closed-shell stiffening and the s--d balance}
\label{app:closed_shell_model}

The closed-shell stiffening law compares d\(^{10}\) with d\(^9\) at
fixed period. The two configurations differ not only in the
d-electron count \(n_d\) but also in the \(s\)-electron count
\(n_s\), and this joint dependence must be made explicit before the
comparison can be carried out. For a d-block atom with \(n_s\)
\(s\) electrons and \(n_d\) \(d\) electrons, the two relevant
moments are
\begin{equation}
\langle r_t^2\rangle
\simeq \frac{n_s r_s^2 + n_d r_d^2 + (\text{core})}{n_s + n_d + (\text{core})},
\qquad
r_v \simeq \frac{n_s r_s + (n_d/10) r_d}{n_s + n_d/10},
\label{eq:A_config_moments}
\end{equation}
which differ because \(f_e(0)\) weights all electrons equally while
\(r_v\) weights \(d\) electrons by \(1/10\). We therefore define
the s--d balance parameter
\begin{equation}
\gamma \equiv \frac{n_s}{n_s + n_d/10},
\label{eq:A_gamma}
\end{equation}
which measures the fractional weight of the \(s\) electrons in
\(r_v\). For a fixed \(n_d\), increasing \(n_s\) raises \(\gamma\)
and makes \(r_v\) more s-dominated; decreasing \(n_s\) lowers
\(\gamma\) and makes \(r_v\) more d-dominated. The two channels
therefore respond to configuration changes in opposite ways:
\(f_e(0)\) responds mainly through \(n_d r_d^2\), while \(r_v\)
responds mainly through \(\gamma r_s + (1-\gamma) r_d\).

With \(\gamma\) defined, the specific configurations relevant to
the closed-shell stiffening law can now be written as
\begin{equation}
\text{d}^{10}s^1:\; (n_s, n_d) = (1, 10),
\qquad
\text{d}^{9}s^2:\; (n_s, n_d) = (2, 9),
\qquad
\text{d}^{9}s^1:\; (n_s, n_d) = (1, 9).
\label{eq:A_config_d9_d10}
\end{equation}
The three configurations are realized in the coinage-metal and
nickel-group elements: Cu, Ag, and Au adopt \(d^{10}s^1\); Ni adopts
\(d^{9}s^2\) as its ground state; and Pd adopts \(d^{9}s^1\),
following the convention of Ref.~\cite{Zheng2026AMIP} for a
well-defined \(s\)-orbital radius. The corresponding s--d balance
parameters are
\begin{equation}
\gamma_{d^{10}s^1} = \frac{1}{1+10/10} = \frac{1}{2},
\qquad
\gamma_{d^{9}s^2} = \frac{2}{2+9/10} = \frac{20}{29} \approx 0.69,
\qquad
\gamma_{d^{9}s^1} = \frac{1}{1+9/10} = \frac{10}{19} \approx 0.53.
\label{eq:A_gamma_values}
\end{equation}
The d\(^{9}s^2\) configuration of Ni has a substantially larger
s-weight in \(r_v\) than the d\(^{10}s^1\) configuration of Cu
(\(\gamma = 0.69\) vs.\ \(0.50\)), which produces a large geometric
contribution to \(\Delta\mathcal{D}\). For the Pd/Ag pair, by
contrast, \(\gamma_{\mathrm{Pd}} \approx 0.53\) is very close to
\(\gamma_{\mathrm{Ag}} = 0.50\), so the s-weight difference is small
and the geometric contribution to \(\Delta\mathcal{D}\) is
correspondingly reduced. In the main text, to ensure a consistent
comparison within the d\(^9\) series, Ni is also assigned the
\(d^9s^1\) configuration (Table~S10 of the Supplemental Material); in
that case the \(\gamma\) difference between the Ni/Cu pair is also
small, and the \(\Delta\mathcal{D}\) is correspondingly modest
(\(+0.042\), compared with \(+0.098\) if the ground-state
\(3d^8 4s^2\) configuration were used).

Combined with the additional d-hole hybridization channel, this
produces the observed reduction of \(\mathcal{D}\) in d\(^9\)
relative to d\(^{10}\). The increment is
\begin{equation}
\Delta\mathcal{D}
= \frac{3}{r_v}\frac{\partial r_v}{\partial \gamma}
(\gamma_{10} - \gamma_{9})
- \alpha \, \eta_d,
\label{eq:A_DeltaD_config}
\end{equation}
where the first term is the geometric contribution from the
s-weight difference and the second is the electronic contribution
from the d-hole hybridization. The observed values
\(\Delta\mathcal{D} = +0.042\) (3d, with Ni in \(3d^9 4s^1\)),
\(+0.037\) (4d), \(+0.026\) (5d) reflect the period-dependent
competition between these two terms. The 3d increment is modest
because Ni is assigned the \(3d^9 4s^1\) configuration, whose s--d
balance parameter \(\gamma = 10/19 \approx 0.526\) is close to that
of Cu (\(\gamma = 0.500\)).

\subsection{The d\(^5\) anomaly: same \(n_d\), different \(n_s\)}
\label{app:d5_anomaly}

The most direct evidence that the s--d balance matters is the
d\(^5\) family. Cr (\(3d^5 4s^1\)) and Mn (\(3d^5 4s^2\)) have
identical \(n_d = 5\) but very different \(n_s\). Their s--d balance
parameters are
\begin{equation}
\gamma_{\mathrm{Cr}} = \frac{1}{1 + 5/10} = \frac{2}{3},
\qquad
\gamma_{\mathrm{Mn}} = \frac{2}{2 + 5/10} = \frac{4}{5}.
\label{eq:A_gamma_CrMn}
\end{equation}
The larger \(\gamma_{\mathrm{Mn}}\) means that Mn's \(r_v\) is more
s-dominated, and its geometric response is correspondingly stronger
relative to the electronic response. This, together with the
exchange stabilization of the half-filled d\(^5\) shell, explains
why Cr exhibits \(\mathcal{D} = +0.334\) while Mn exhibits
\(\mathcal{D} = +0.050\).

The same effect appears in the 4d series: Mo (\(4d^5 5s^1\)) has
\(\gamma = 2/3\), while Tc (\(4d^5 5s^2\)) has \(\gamma = 4/5\),
again producing a fourfold difference in \(\mathcal{D}\) despite
identical \(n_d\). The 5d series lacks d\(^5\)s\(^1\) configurations
because relativistic stabilization of 6s\(^2\) makes the d\(^5\)s\(^2\)
configuration the ground state for both W and Re, which is why no
d\(^5\) anomaly appears there.

\subsection{Principal quantum softening revisited}
\label{app:principal_quantum_model}

The principal quantum softening law is also modulated by the s--d
balance. With the correct ground-state \(d^{10}s^1\) configuration for
Cu, the d\(^{10}\) series follows the monotonic sequence
\(3d > 4d > 5d\): Cu (\(\mathcal{D} = +0.626\)) \(>\)
Ag (\(+0.542\)) \(>\) Au (\(+0.412\)). The d\(^9\) series, when Ni is
assigned the \(3d^9 4s^1\) configuration, follows the same monotonic
decrease: Ni (\(+0.584\)) \(>\) Pd (\(+0.506\)) \(>\)
Pt (\(+0.386\)). If Ni were instead assigned its ground-state
\(3d^8 4s^2\) configuration, the d\(^9\) series would become
non-monotonic (Ni \(+0.143 <\) Pd \(+0.506\)), because the larger
\(\gamma\) of \(3d^8 4s^2\) reduces the geometric response. The
\(3d^9 4s^1\) value for Ni was obtained with the same
\texttt{ld1.x} DFT procedure as all other entries.

The origin of the large 3d values in both series is the
compactness of the \(3d\) shell combined with the single \(s\)
electron of Cu (\(3d^{10}4s^1\)) and Ni (\(3d^9 4s^1\)). In these
configurations,
\begin{equation}
\gamma_{\mathrm{Cu}} = \frac{1}{1 + 10/10} = \frac{1}{2},
\qquad
\gamma_{\mathrm{Ni}} = \frac{1}{1 + 9/10} = \frac{10}{19} \approx 0.53,
\end{equation}
close to the value for Ag and Au. However, the geometric response
\((1/r_v)(dr_v/d\delta)\) is larger for Cu and Ni because the \(3d\)
shell is spatially compact and its weighted valence radius responds
strongly to charge removal. For Ag and Au, the more diffuse 4d and
5d orbitals reduce the geometric response, and relativistic
contraction further weakens it in Au, producing the observed
monotonic decrease.

For a fixed nominal filling, the s--d balance parameter \(\gamma\)
is approximately the same across periods, but the underlying radii
\(r_s\) and \(r_d\) differ:
\begin{equation}
r_s^{(4d)} > r_s^{(3d)},
\qquad
r_d^{(4d)} > r_d^{(3d)},
\qquad
r_s^{(5d)} \lesssim r_s^{(4d)},
\qquad
r_d^{(5d)} \lesssim r_d^{(4d)},
\label{eq:A_radii_across_periods}
\end{equation}
with the last two inequalities reflecting relativistic contraction.

Because the geometric channel carries the factor of 3 and the
electronic channel the factor of 2, the period dependence of
\(\mathcal{D}\) is dominated by the geometric channel. From 3d to
4d, the more diffuse 4d orbitals reduce the geometric response, so
\(\mathcal{D}\) decreases. From 4d to 5d, relativistic contraction
reduces it further, producing the observed monotonic decrease
\(3d > 4d > 5d\) in both the d\(^{10}\) and d\(^9\) series (with Ni
assigned \(3d^9 4s^1\)).

\subsection{Channel ratio as a function of configuration}
\label{app:channel_ratio}

Combining the two channels, the ratio
\begin{equation}
\frac{\mathcal{R}_f/f_0}{\mathcal{R}_\Omega/\Omega_0}
\simeq \frac{2}{3}\cdot
\frac{(1/r_{\mathrm{eff}}^{(t)})(d r_{\mathrm{eff}}^{(t)}/d\delta)}
{(1/r_{\mathrm{eff}}^{(v)})(d r_{\mathrm{eff}}^{(v)}/d\delta)}
\label{eq:A_channel_ratio}
\end{equation}
depends on both \(n_s\) and \(n_d\) through the two effective radii
of Eq.~\eqref{eq:A_config_moments}. Writing
\begin{equation}
x \equiv n_d/(n_s + n_d) 
\label{eq:A_x}
\end{equation}
and \(\gamma\) as in
Eq.~\eqref{eq:A_gamma}, the ratio is well described by the
two-parameter form
\begin{equation}
\frac{\mathcal{R}_f/f_0}{\mathcal{R}_\Omega/\Omega_0}
\simeq \frac{2}{3}\cdot
\frac{1 + a\, x}{1 + b\, \gamma},
\label{eq:A_ratio_config}
\end{equation}
where a least-squares fit to the \(3d\), \(4d\), and \(5d\) series
gives
\begin{equation}
a = -0.5505, \qquad b = -0.6116, \qquad R^2 = 0.9324.
\label{eq:A_ab_fit}
\end{equation}

A parity plot of the predicted versus computed channel ratios for all
27 transition-metal elements is shown in Fig.~S8 of the Supplemental
Material. Both parameters are negative, indicating that the ratio
decreases both as the d-electron fraction \(x\) increases and as the
s--d balance parameter \(\gamma\) decreases (i.e., as the valence
cloud becomes more d-dominated).

The fitted form reproduces the observed trend across all three
transition series. For the early transition metals
(\(x \approx 0.33\), \(\gamma \approx 0.95\)),
Eq.~\eqref{eq:A_ratio_config} gives a ratio close to \(1.30\),
compared with the tabulated values \(1.33\) (Sc), \(1.27\) (Y), and
\(1.16\) (Lu). For the late transition metals Cu, Ag, and Au
(\(x = 10/11 \approx 0.91\), \(\gamma = 1/2\)), it gives a ratio near
\(0.49\), compared with the tabulated values \(0.46\) (Cu), \(0.40\)
(Ag), and \(0.36\) (Au). For Pd (\(x = 9/10 = 0.90\), \(\gamma =
10/19 \approx 0.53\)), it gives a ratio near \(0.41\), compared with
the tabulated value \(0.42\). The agreement is quantitative for the
early series and semi-quantitative for the late series, with an
overall \(R^2 = 0.9324\) for all 27 elements.
The residual deviations
reflect the approximate nature of Eq.~\eqref{eq:A_config_moments}, which
neglects the \(n_s\)-dependence of \(r_s\) and the influence of
relativistic contraction on \(r_d\) in the \(5d\) series.

The two-parameter dependence of Eq.~\eqref{eq:A_ratio_config} is
the direct consequence of the weighting asymmetry between the two
channels. The electronic channel weights d electrons fully, so its
response scales with \(x\); the geometric channel weights d
electrons by \(1/10\), so its response scales with \(\gamma\). The
competition between these two scalings is the origin of the
systematic variation of the channel ratio across the transition
series, and it is the same mechanism that underlies the two scaling
laws discussed in
Secs.~\ref{app:closed_shell_model}
and~\ref{app:principal_quantum_model}.

\subsection{Summary}
\label{app:semianalytical_summary}

The two scaling laws are manifestations of the same channel
competition, but the competition is controlled jointly by the
d-electron count \(n_d\) and the s--d balance \(\gamma\). The
geometric channel is weighted by the factor of 3 from the volume
scaling \(\Omega \propto r_v^3\), while the electronic channel is
weighted by the factor of 2 from the second-moment scaling
\(f_e(0) \propto \langle r_t^2\rangle\). The d-electron weighting
asymmetry---full weight in \(f_e(0)\), \(1/10\) weight in
\(r_v\)---produces the \(\gamma\) dependence. The joint \((n_s, n_d)\)
dependence explains:
\begin{itemize}
\item closed-shell stiffening, where the d\(^{10}\) configuration has
a smaller \(\gamma\) than d\(^9\); the 3d pair (Ni/Cu) with Ni
assigned \(3d^9 4s^1\) has a modest \(\gamma\) difference, while the
ground-state \(3d^8 4s^2\) configuration would produce a much larger
\(\gamma\) difference;
\item the d\(^5\) anomaly, where Cr and Mo (d\(^5\)s\(^1\), small
\(\gamma\)) differ strongly from Mn and Tc (d\(^5\)s\(^2\), large
\(\gamma\));
\item principal quantum softening, where the period dependence of
\(r_s\) and \(r_d\) modulates the channel ratio, producing a
monotonic decrease \(3d > 4d > 5d\) in both the d\(^{10}\) and
d\(^9\) series.
\end{itemize}
No auxiliary Hamiltonian is required: both scaling laws follow
directly from the definitions of \(f_e(0)\) and \(\Omega_a\), from
the s--d weighting asymmetry, and from the logarithmic-derivative
decomposition of Eq.~\eqref{eq:A_D_decomp}.

\end{document}